\documentclass{aa}
\usepackage{psfig}
\usepackage[dvips]{graphicx}
\usepackage[]{longtable}

\newcommand{\ogle}[1]{\object{OGLE-TR-{#1}}}

\def\beq{\begin{equation}}
\def\eeq{\end{equation}}
\def\av{$A_{\mathrm{V}}$}
\def\rv{$R_{\mathrm{V}}$}
\def\ebv{$E(B-V)$}

\begin{document}

\title{Characterisation of extrasolar planetary transit candidates 
       \thanks{Based on observations collected at the European
        Southern Observatory, La Silla, Chile, within the Observing 
        Programme 70.B-0547(B).}
       }

\author{
J. Gallardo \inst{1,2}, 
D. Minniti \inst{2}, 
D. Valls-Gabaud \inst{3,4}
and
M. Rejkuba \inst{5}
}

\offprints{J. Gallardo, \email{jose.gallardo@ens-lyon.fr}}

\institute{CNRS UMR 5574, C.R.A.L., \'Ecole Normale Sup\'erieure, 69364 Lyon Cedex 07, France 
     \and
Departamento de Astronom\'{\i}a y Astrof\'{\i}sica, P. Universidad Cat\'olica, 
                         Av. Vicu\~na Mackenna 4860, Santiago, Chile 
     \and
Canada-France-Hawaii Telescope, 65-1238 Mamalahoa Highway, Kamuela, Hawaii 96743, USA 
     \and 
CNRS UMR 5572, Laboratoire d'Astrophysique, Observatoire Midi-Pyr\'en\'ees,
               14 Avenue \'Edouard Belin, 31400 Toulouse, France
     \and
European Southern Observatory, Karl-Schwarzschild-Stra{\ss}e 2, 85748 Garching, Germany 
          }

\date{Received ... / Accepted ... }

\titlerunning{Extrasolar planetary transit candidates}
\authorrunning{Gallardo et al.}

\abstract{The detection of transits is an efficient technique to uncover faint companions 
around stars. The full characterisation of the companions (M-type stars, brown dwarfs or exoplanets) 
requires high-resolution spectroscopy to measure properly masses and radii. With the advent of massive
variability surveys over wide fields, the large number of possible candidates makes such a full
characterisation impractical for all of them. We explore here a fast technique to pre-select the 
most promising candidates
using either near-IR photometry or low resolution spectroscopy. We develop a new method
based on the well-calibrated surface brightness relation along with the correlation between mass 
and luminosity for main sequence stars, so that not only giant stars can be excluded but also 
accurate effective temperatures and radii can be measured. The main source of uncertainty arises from the
unknown dispersion of extinction at a given distance. We apply this technique to our observations
of a sample of 34 stars extracted from the 62 low-depth transits identified by OGLE during their survey of 
some 10$^5$ stars in the Carina fields of the Galactic disc. We infer that at least 78\% of the companions
of the stars which are well characterised in this sample are not exoplanets. Stars \ogle{105}, 
\ogle{109} and \ogle{111} are the likeliest to host exoplanets and deserve high-resolution 
follow-up studies. \ogle{111} was very recently confirmed as an exoplanet with M$_{planet}$ $\cong$ 0.53 $\pm$ 0.11 M$_{Jup}$ (Pont et al. 2004), confirming the efficiency of our method in pre-selecting reliable planetary transit candidates.  
}%

\maketitle

\section{Introduction}
%---------------------

The search for extra-solar planets (exoplanets) is becoming increasingly important, yielding
new insights both in stellar and planetary physics (e.g. Gonzalez 2003). Whilst 
some studies (Lineweaver \& Grether 2003) suggest that 100\% of Sun-like stars may have
planets, current surveys  appear to be heavily biased by the detection techniques used. 
The radial velocity technique has proved to be extremely successful over the past  decade  
in discovering  exoplanets, even though few planetary characteristics can be derived
from the measurements. In contrast, planetary transits may yield many more properties, provided
the host stars are well characterised. While transit surveys may monitor many orders of
magnitude more stars in a comparatively shorter observing time than radial velocity surveys, they
are not  devoid of problems. Blending, grazing eclipses and contamination by giant stars, among
other problems, make some
detections somewhat ambiguous (Dreizler et al. 2002, Brown et al. 2001, Drake 2003, Drake \& 
Cook 2004). Clearly 
transit surveys have the ability to discover many interesting exoplanet candidates, but 
non-planetary events must be filtered out. The best way to perform such a separation is to 
obtain 
radial velocities at different phases along the orbit. Whilst this approach is becoming very 
successful
(Dreizler et al. 2002, Konacki et al. 2003, Bruntt et al. 2003, Street et al. 2003, Bouchy et 
al. 2004), 
precision radial velocities for all transit candidates, 
especially those at faint magnitudes, would be far too time consuming, and so pre-selections
 must  be
carried out. For instance, Dreizler et al. (2002) made a selection of the best planetary 
candidates 
by using low dispersion spectroscopy to refine stellar parameters, while Drake  (2003) used the 
gravity darkening dependence of the light curves. Another possibility is  near-IR photometry
which can also be used to discard giant stars, stars with IR excess, and to better determine 
stellar parameters (e.g. see Ribas et al. 2003, for the case of bright stars with 2MASS 
photometry) and 
thus the radii of the companion objects. Ellipsoidal variability is also a good way to filter
out the larger companions, although the correlation in the photometric time series limits in
practice its applicability (Sirko \& Paczy\'nsky 2003).

The characterisation of the companions and their host stars is thus important for two main 
reasons. First, it allows a pre-selection of the most promising exoplanet candidates, and 
filters out giant stars and large companions. 
Second, the depth of the transits depends upon $R_c/R_*$, the ratio of the radius of the companion
$R_c$ to the one from the host star $R_*$. Yet the most important and interesting quantity
in the case of exoplanets is the radius $R_c$, as it allows inferences on the planet's composition
and evolutionary history. Characterising the properties of the host star, and in particular its
radius $R_*$ is therefore very important for transit candidates. 

In this work we explore the possibility to characterise, through IR photometry and low dispersion
 spectroscopy, the properties of 
low-luminosity stars whose transits were discovered by the OGLE-III collaboration (Udalski 
et al. 2002b),
in order to constrain astrophysical parameters of both stellar and companion objects, some of
which may be exoplanets.

In Sections 2 and 3 we present the observations and data reduction. We describe the estimation 
of stellar parameters in Section 4 through a novel application of the surface brightness method. 
The inferred properties of the companions are discussed in Section 5. A brief discussion of the
stellar populations observed in these fields is provided in Section 6 with the help of a model
of the Galaxy. 

\section{Observations}
%---------------------

In 2002 the second ``planetary and low luminosity'' companion catalogue was published by the 
Optical Gravitational Lensing Experiment (OGLE). In this campaign, the OGLE team observed
 3 fields 
in the Carina region of the Galactic disc. These fields were monitored in February--May 2002 
searching 
for low-depth transits in about 100000 stars, and 62 objects were discovered with transit
depths shallower than 0.08 mag in the $I$ band. Many of these objects may be extrasolar planets, 
brown dwarfs or 
M-type stars (see Udalski et al. 2002b for further details). In order to characterise both
stars and companions, we carried out IR imaging and optical spectroscopy in the two Carina 
fields (Table~\ref{table:fields}).
Table~\ref{table:logobs} lists the schedule of observations for the subset of planetary transit 
candidates analysed in this paper.

\subsection{Optical $VI$ imaging from OGLE}

The optical photometry in the $V$ and $I$ bands come from the  catalogue by OGLE (Udalski et 
al. 2002b), whose
observations were carried out with the 1.3m Warsaw telescope at the Las Campanas Observatory,
 Chile
(operated by the Carnegie Institution of Washington).
The camera used was a wide field CCD mosaic of eight 2048 $\times$ 4096 pixel SITe ST002A 
detectors, with
a field of view of about 35\arcmin $\times$35\arcmin~ giving a scale of 0\farcs26 per pixel. 
The data were obtained during 76 nights starting from February 17, 2002. The fields were
 monitored 
$\sim$ 6 hours per night in the I-band filter with 180 seconds exposure time for each image.
 The median 
seeing for the all data was about 1\farcs2. The names and coordinates for the observed fields 
are 
provided in Table~\ref{table:fields}.

\subsection{$K$-band imaging}

We have acquired photometric data in the $K$ band  for the CAR 104 and CAR 105 fields. 
Observations of 
each field covered an area 35\arcmin$\times$45\arcmin~ with the centres listed in 
Table~\ref{table:fields}. 
Exposure 
times were 3 seconds per image. The images were taken typically  at airmass $\approx$ 1.2 and 
the 
limit magnitude was $K \approx $ 17.

These images were taken on January 18, 2003, as a backup program, with the SOFI near-IR array 
at the 
3.5m New Technology Telescope at the ESO La Silla Observatory in Chile. The detector was a 
Hawaii 
HgCdTe 1024 $\times$ 1024 array with a scale of 0\farcs29 per pixel and a field of view of 
5\arcmin
$\times$5\arcmin. 
The gain was $\sim$ 5.6 e$^{-}$/ADU and the readout noise 2.5 ADU.

\subsection{Low resolution spectroscopy}

The spectroscopy was carried out at the Magellan II 6.5m Clay telescope equipped with the 
Low Dispersion 
Spectrograph S2 (LDSS2), during the nights of May 10 and 11, 2003.
The data were acquired as a backup program on bright nights, with some thin cirrus and 
seeing of the order 
of 1 arcsec. We used the 1\arcsec\ wide slit, with the high re\-so\-lu\-tion grating
which gives 6 \AA\  resolution and a spectral coverage from 3700 \AA~ $ < \lambda < $ 7500 \AA.
The spectra were taken at low airmass ($<$1.2), and the exposure times ranged
from 90 sec to 600 sec for these relatively bright stars ($14.4\leq V \leq 18.1$).
A HeNeAr lamp exposure was acquired immediately after each star's spectrum.
We managed to observe a dozen of sources selected from the list of OGLE candidates 
(Table~\ref{table:logobs}).
%We also observed the spectrophotometric standards LTT3218 and LTT4816, and the
%spectral type and radial velocity standards HR3418, HR6136, HR7429, and  giants
%from the globular cluster $\omega$ Cen.

\section{Data reduction}
%-----------------------

\subsection{Optical Photometry from OGLE}

We used the standard, pipeline-reduced data from OGLE-III (second ``planetary and 
low-luminosity object 
transit'' release, see Udalski et al. 2002a, 2002b). The light curves were
produced by a difference image analysis (DIA), and the calibration was done using the OGLE-II 
and OGLE-III data with  errors that should not exceed 0.1--0.15 mag. The very precise DIA
photometry of some 100,000 stars yielded  62 objects with shallow dips in their light curves, 
suggesting flat transits.

\subsection{IR photometry}

\begin{figure*}[h!]
\centering
\includegraphics[height=10cm,width=10cm]{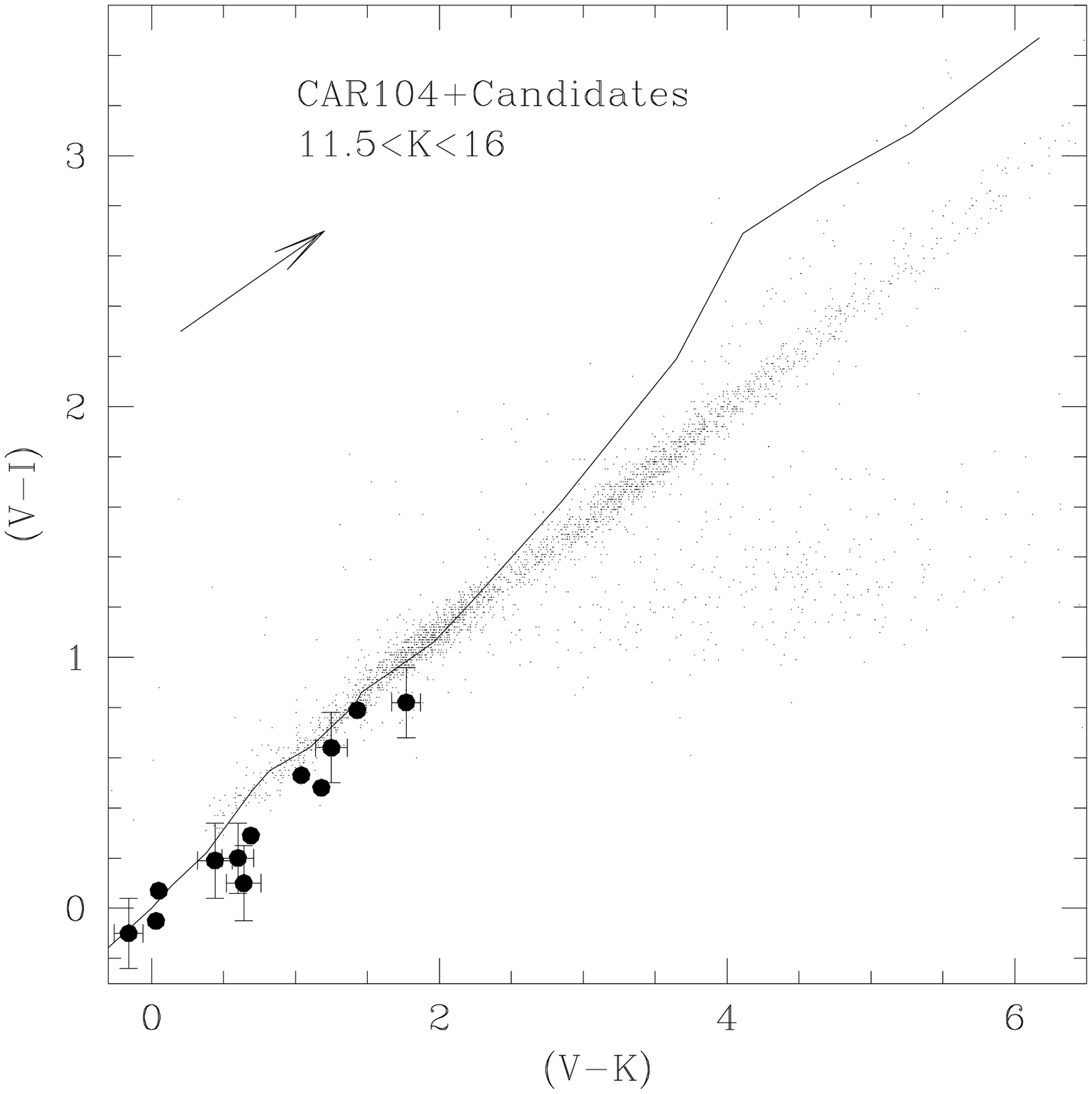}
\includegraphics[height=10cm,width=10cm]{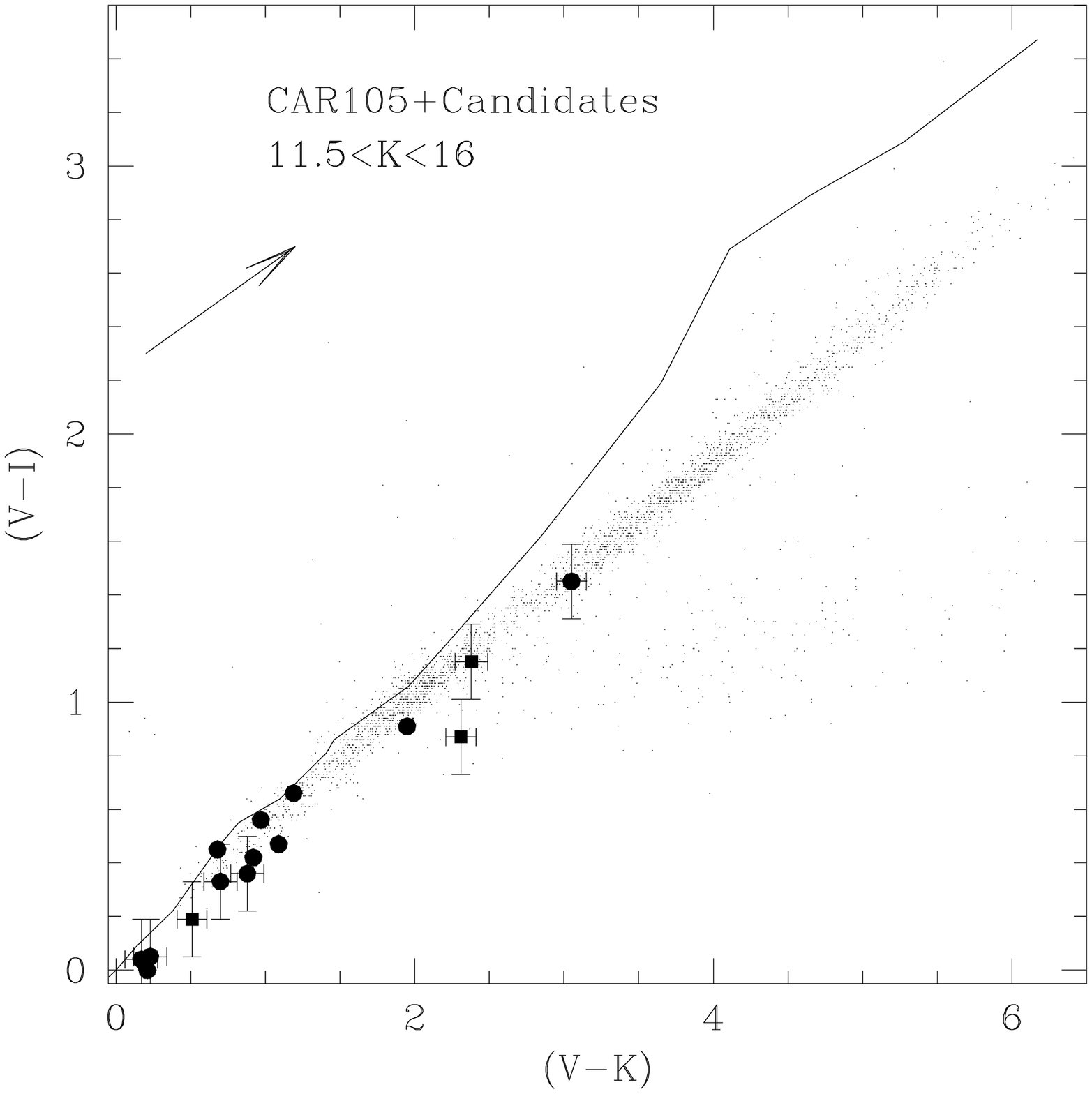}
\caption[]{Colour-colour diagram for CAR 104 and CAR 105 fields. The  circles are the transit 
candidates and 
the solid line represents the locus of main sequence stars. 
The arrows give the direction of the reddening vector. 
Note that, unlike the field stars, stars with transits have been corrected for reddening. 
and that the most likely transit candidates to
host exoplanets are marked with full squares.
}
\label{fig:colcol}
\end{figure*}

The reduction of the $K$-band data included flat-field and sky substraction, which was carried
 out with
IRAF\footnote{IRAF is distributed by the National Astronomy Observatories, which is operated 
by the 
Association of Universities for Research in Astronomy, Inc. (AURA) under cooperative agreement 
with the National Science Foundation}. The sky frame, which was first subtracted from each 
image, was  made 
by median-combining 3 to 4 adjacent images. The subsequent flat-fielding was done with a 
master-flat 
made from all the frames in the mosaic. The instrumental magnitude was calculated with 
Sextractor 
(Bertin \& Arnouts 1996) using a median FWHM of about 2.7 pixels for all the stars. The 
aperture radius 
was 1 FWHM and the background was taken at 2 FWHM. The zero point calibration was done 
using the 2MASS 
(2 Micron All Sky Survey) database. We required a match between the position of our stars 
and the 
2MASS stars within 1\arcsec~ in RA and DEC. Our magnitude range was between $11 < K < 17$
 and the error 
in the zero point calibration was about 0.05 (median for this filter calculated by 2MASS).
 Note that the 
error shown in Table~\ref{table:VIK} 
does not include the zero point error. The $K$-band zero point determined was 
$Z_{p}$= 23.43. For the $V$ magnitude, we matched the published list by Udalski et al. (2002b)
 and a 
list delivered by the OGLE project in which we have the photometric $V$ and $I$ magnitudes
 and coordinates
for all the stars in the two fields. Because the fields are somewhat crowded, some candidates
 were 
surrounded by other stars and the match failed. Table~\ref{table:VIK} summarises the 
photometry for 
the target stars. Column 1 gives the stellar ID from Udalski et al. (2002b), columns 2 and 
3 the optical
$VI$-band photometry from OGLE, and column 4  our $K$-band photometry along with its error. Other
columns are discussed later on. 
The target stars have $14.3 \leq V \leq 18.65$, and $12.17 \leq K \leq 16.4$. The resulting 
optical-infrared colour-colour diagram is given on Figure~\ref{fig:colcol}. 

\subsection{Spectroscopy}

\begin{figure*}[h]
\begin{center}
\leavevmode
\psfig{file=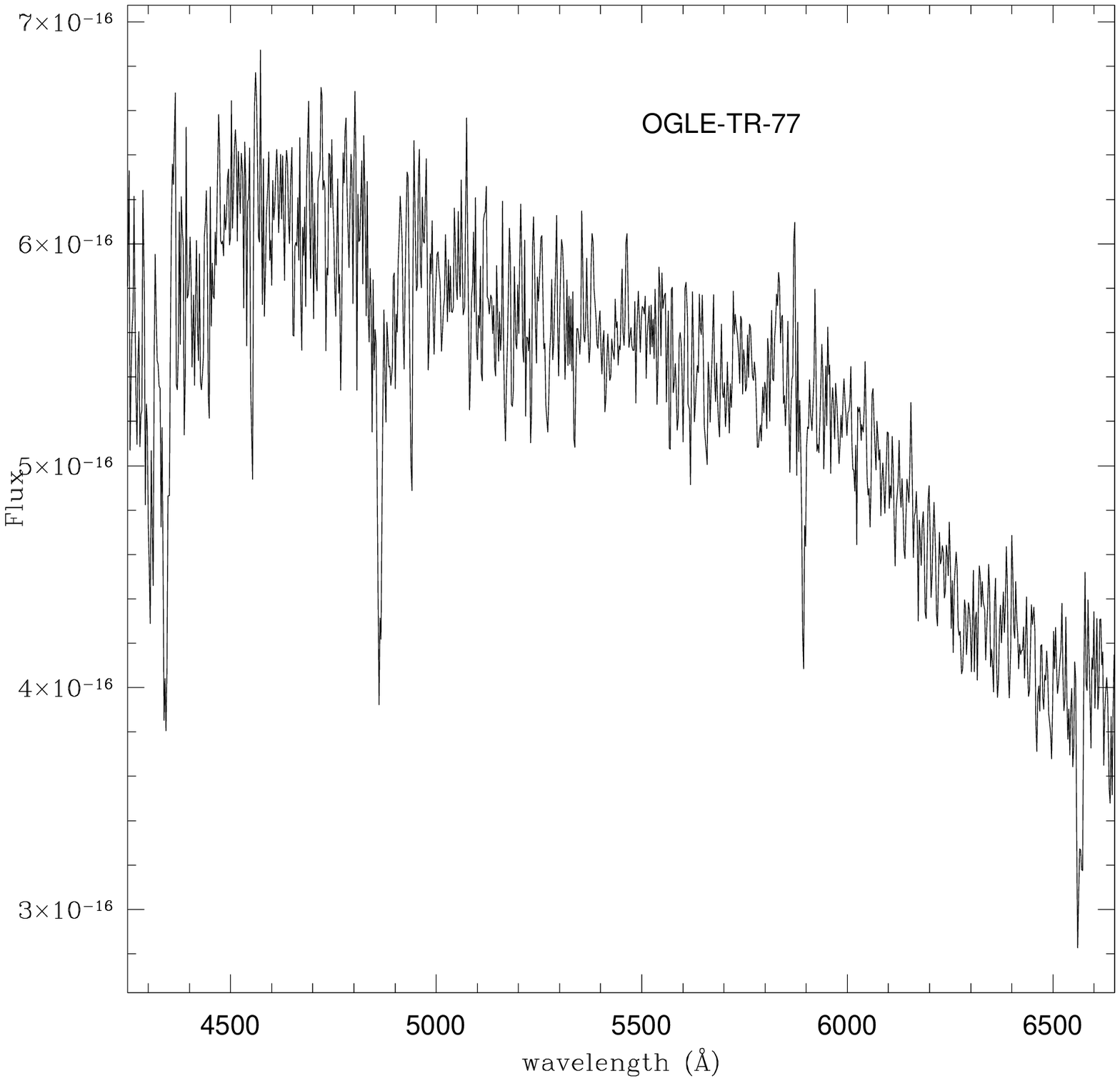,height=50mm}
\psfig{file=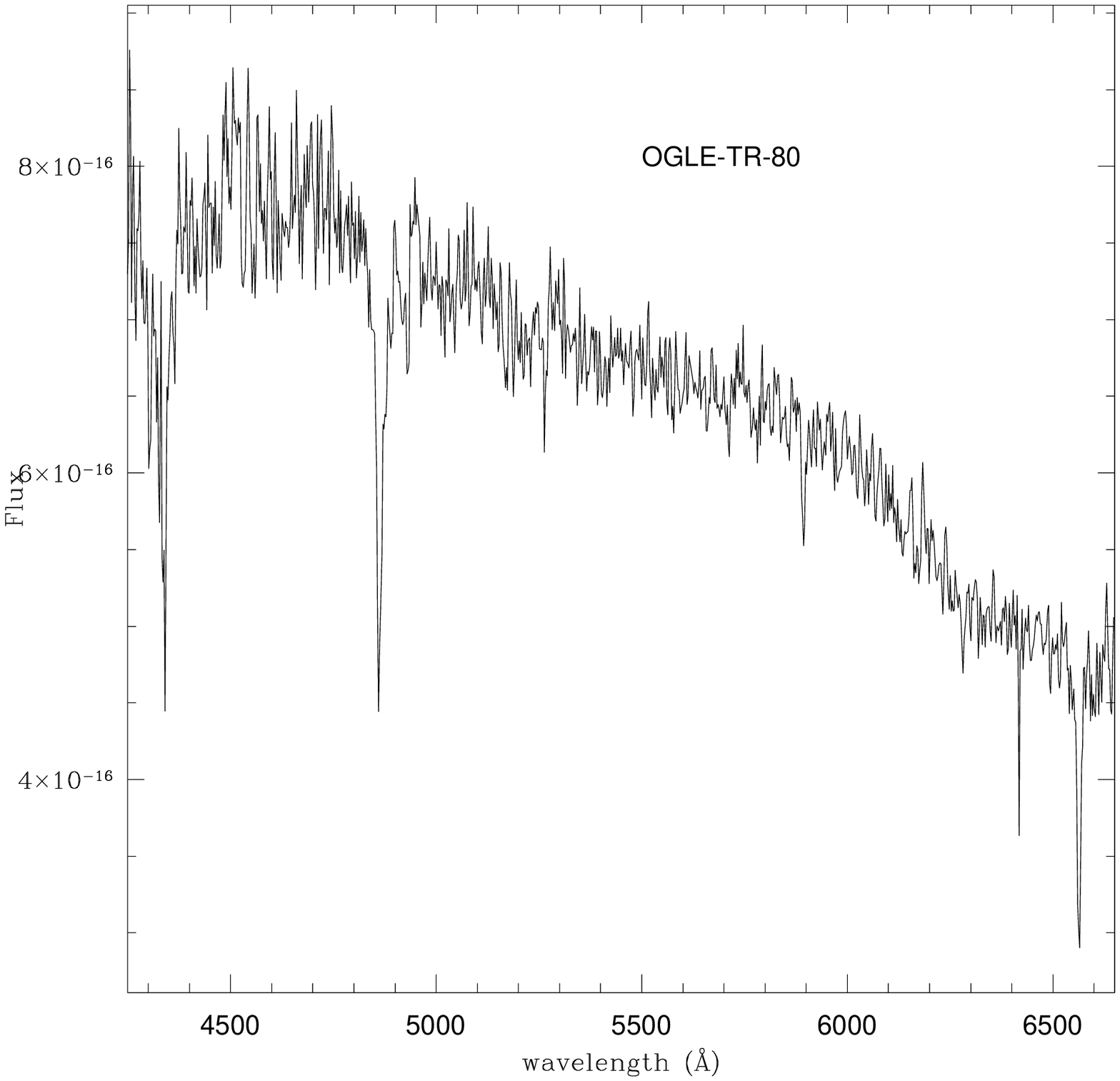,height=50mm}
\psfig{file=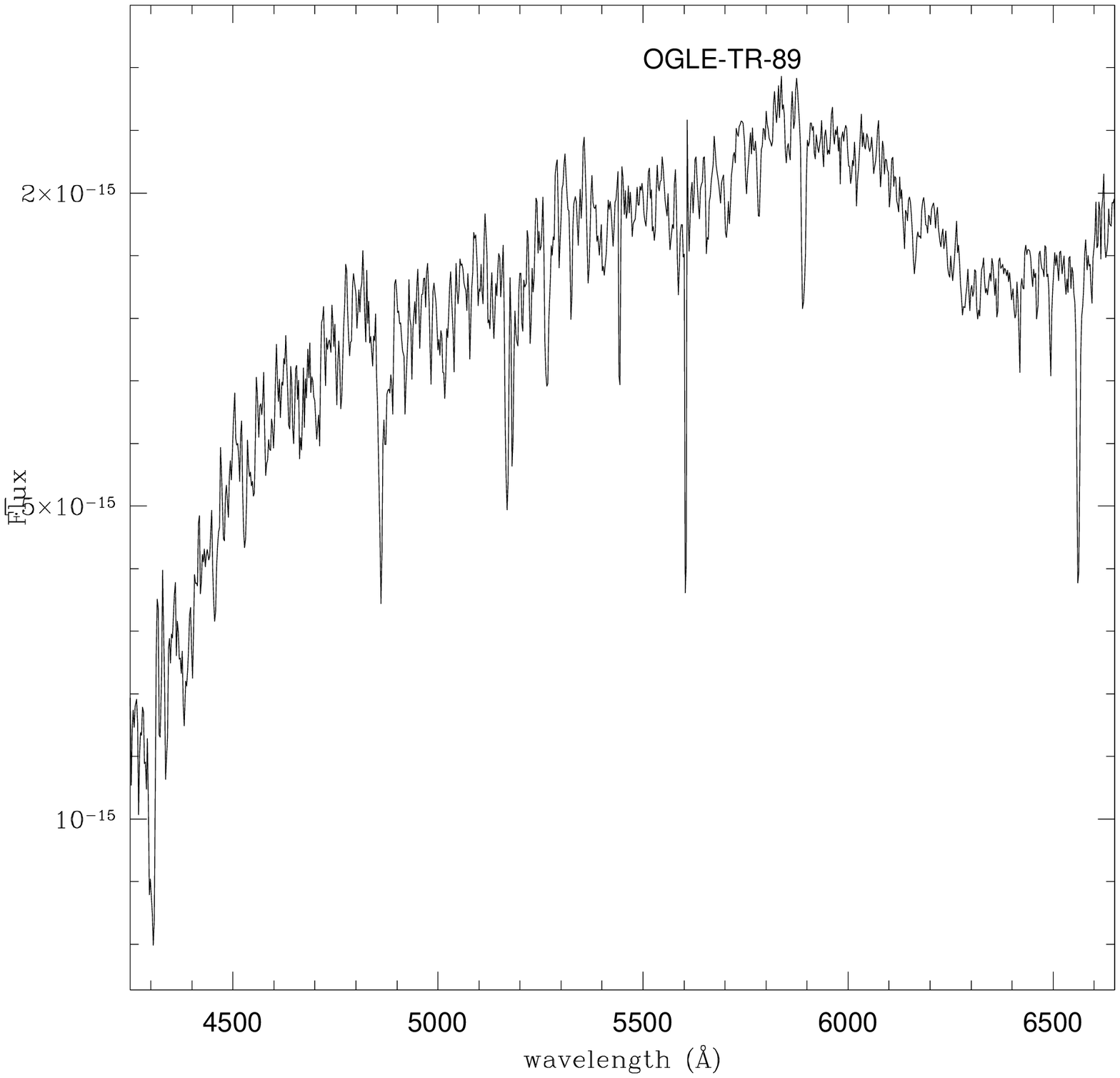,height=50mm}
\psfig{file=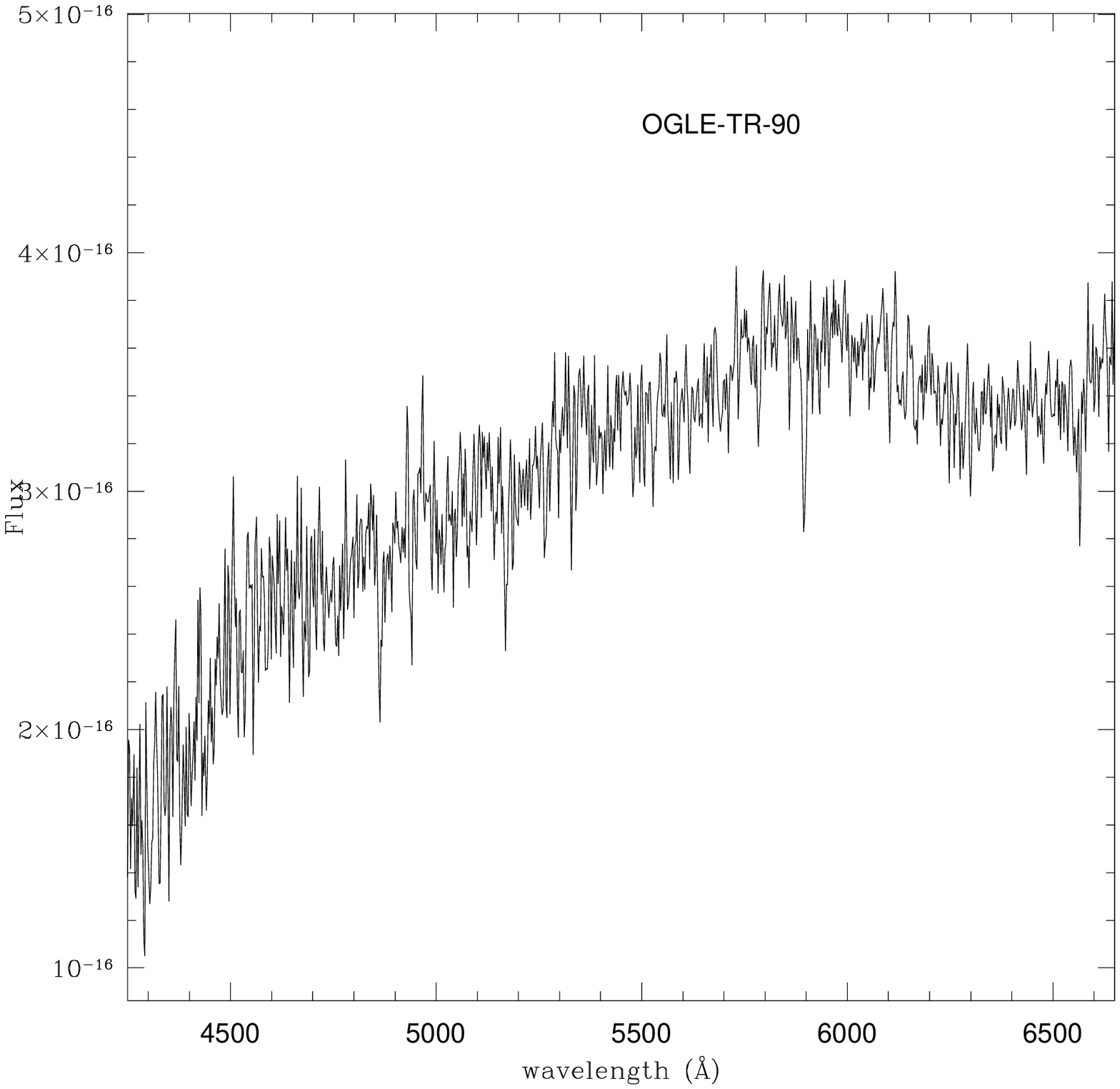,height=50mm}
\psfig{file=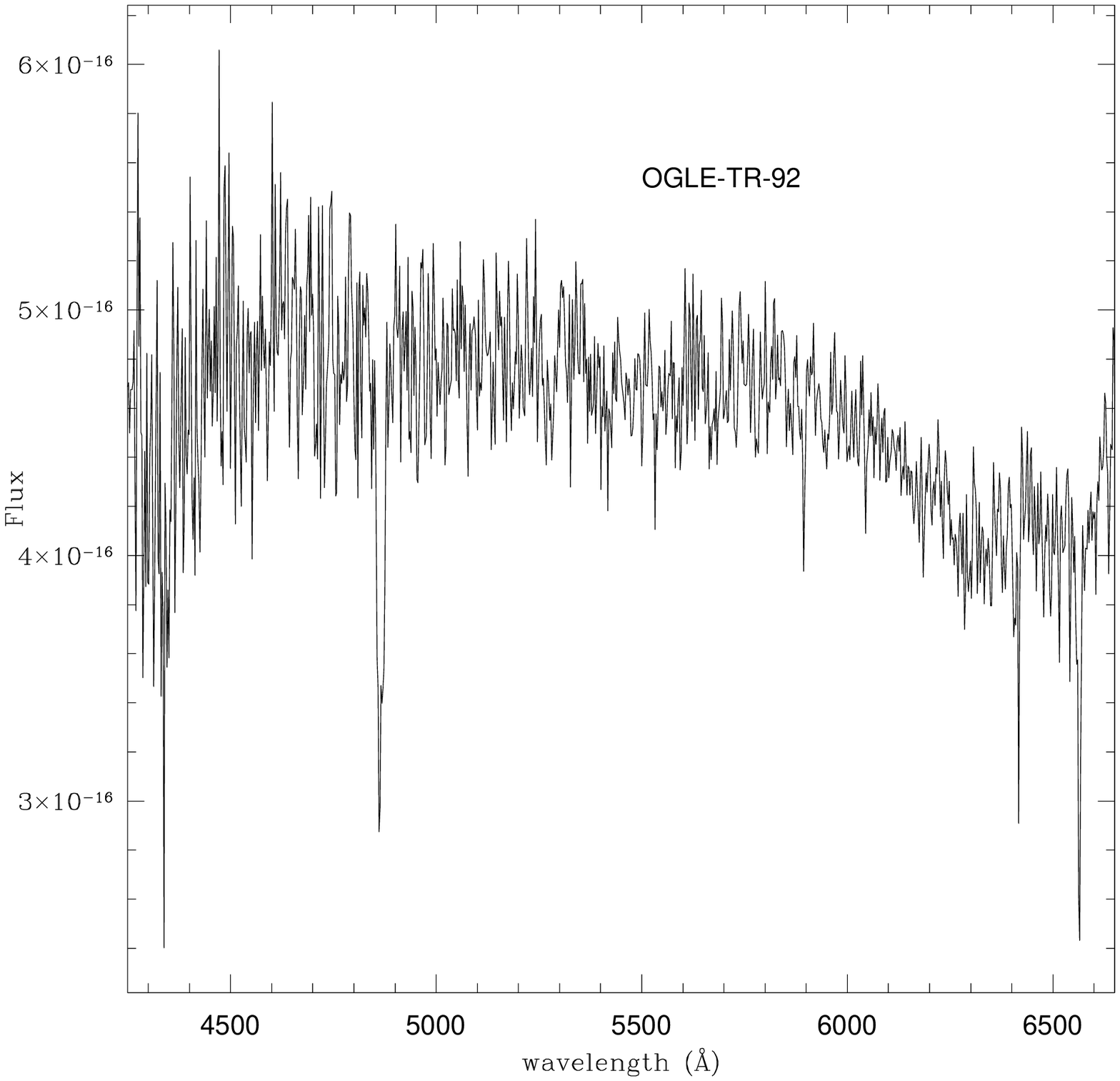,height=50mm}
\psfig{file=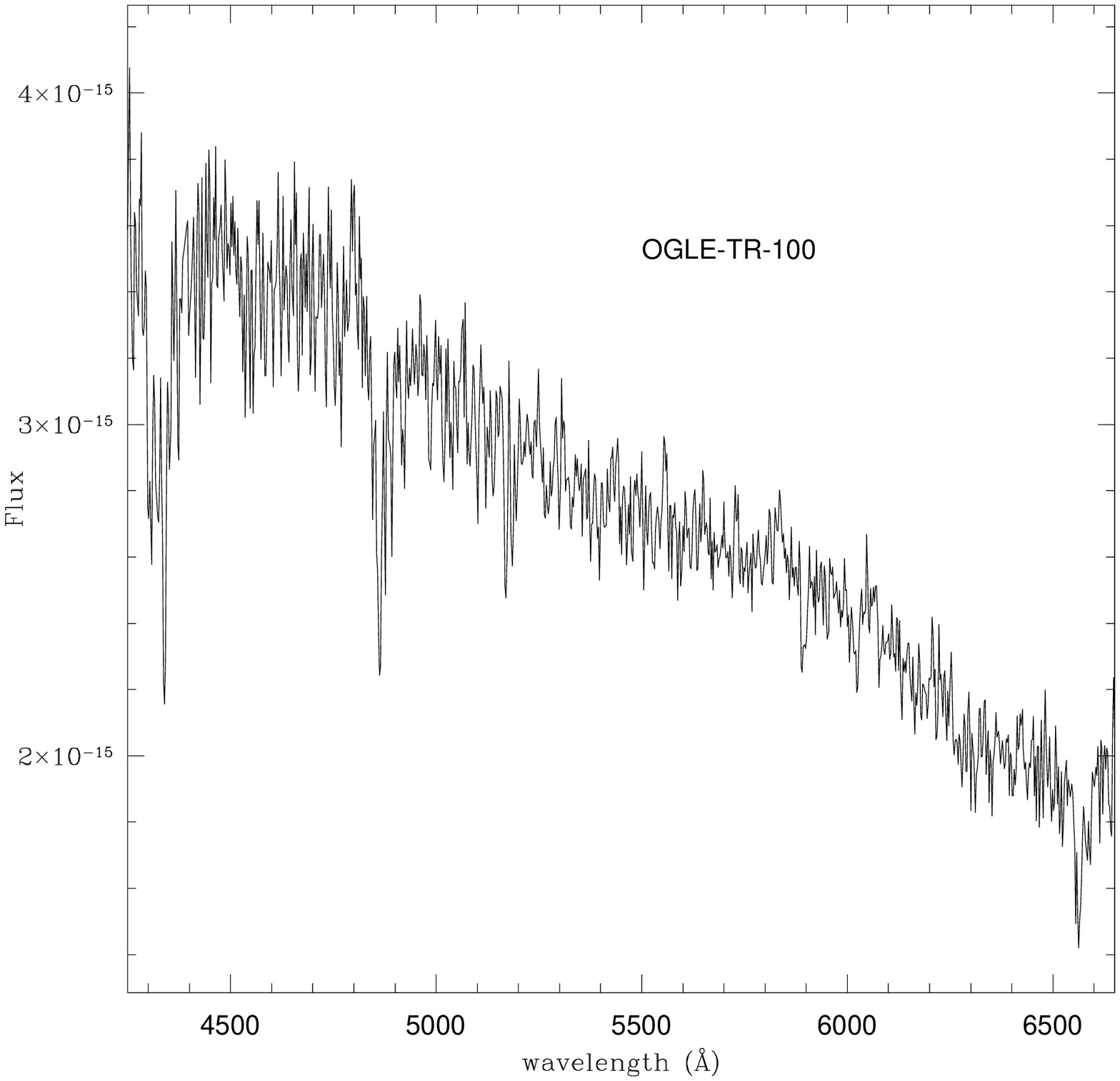,height=50mm}
\psfig{file=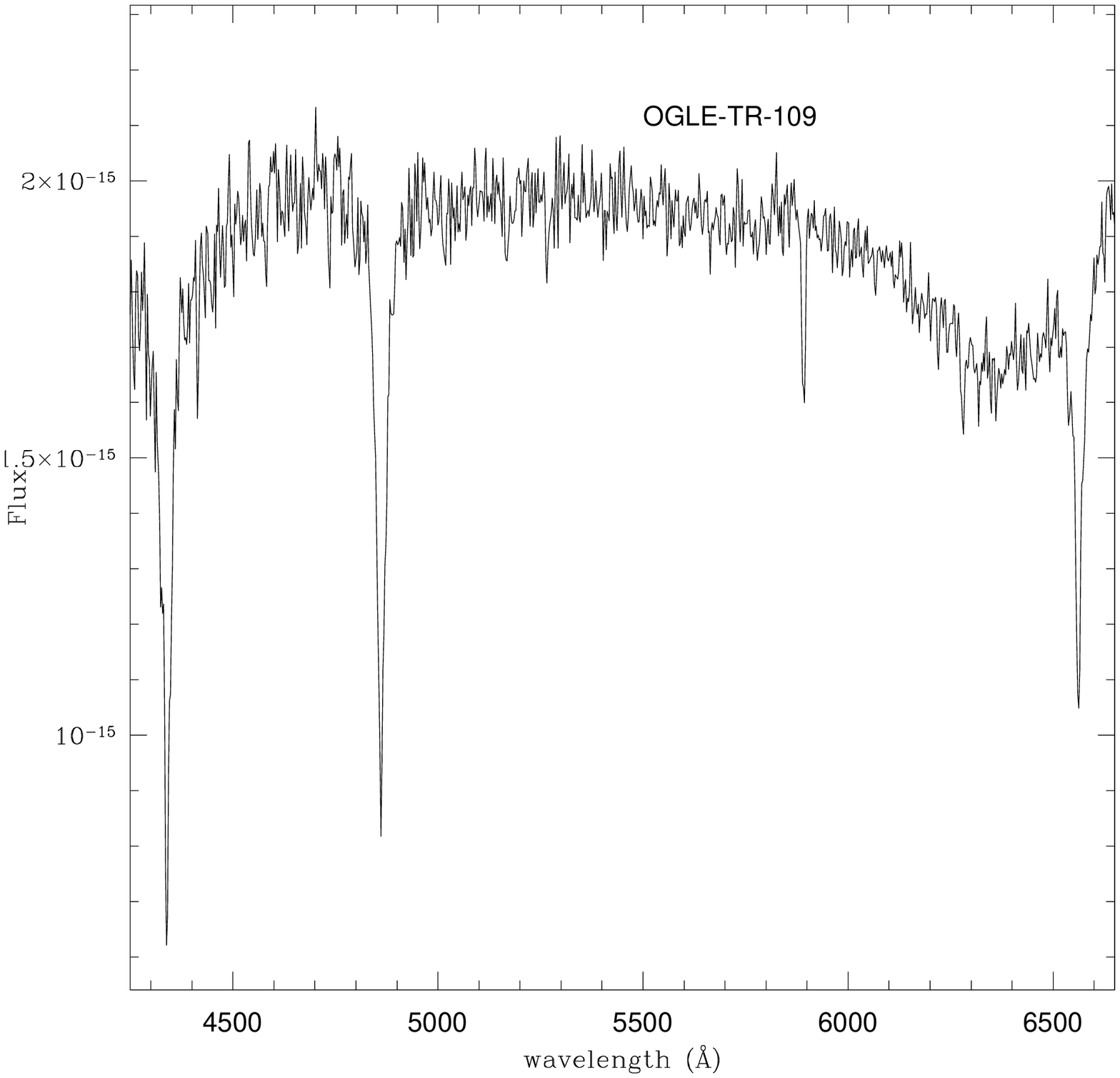,height=50mm}
\psfig{file=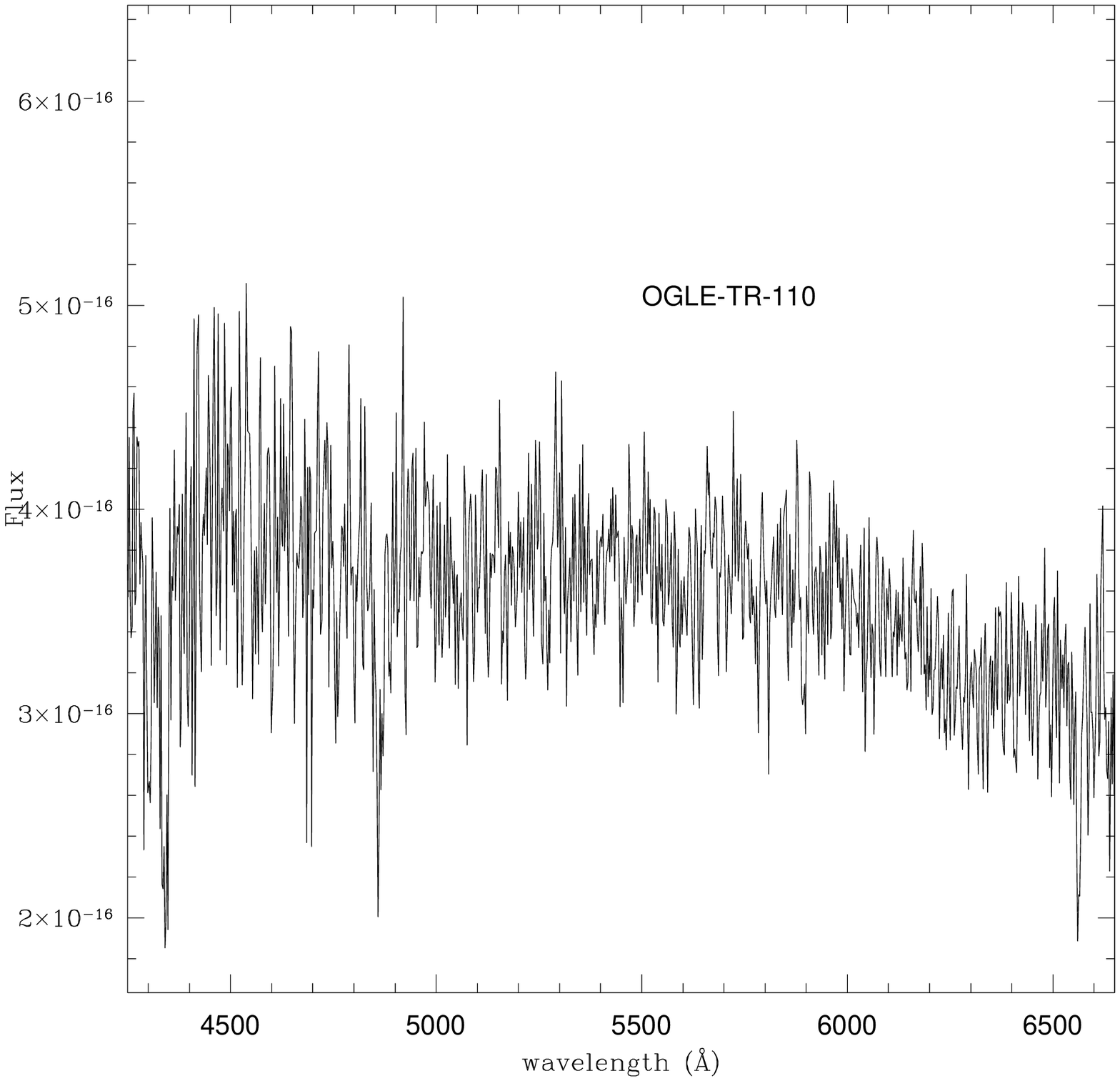,height=50mm}
\psfig{file=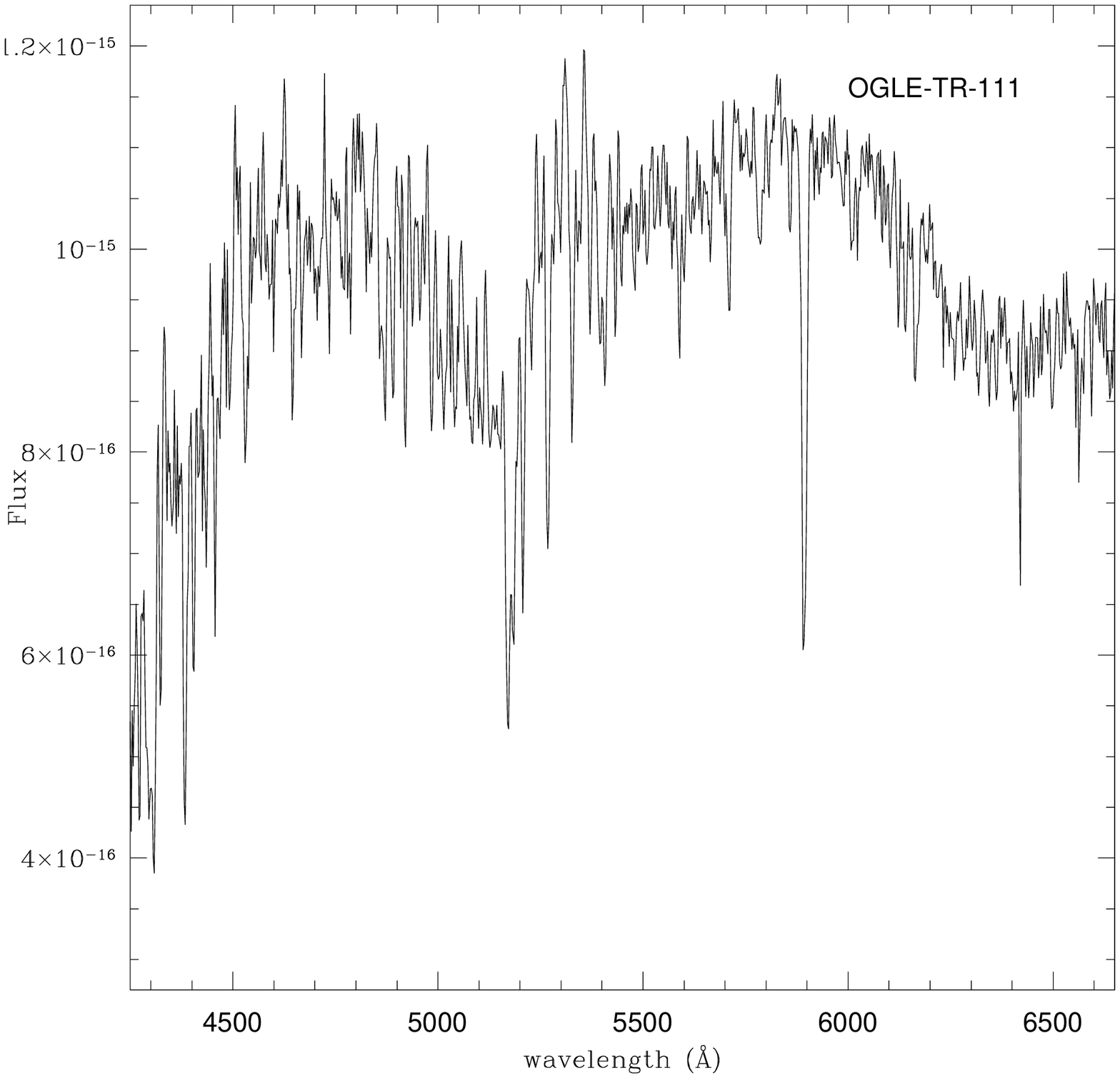,height=50mm}
\psfig{file=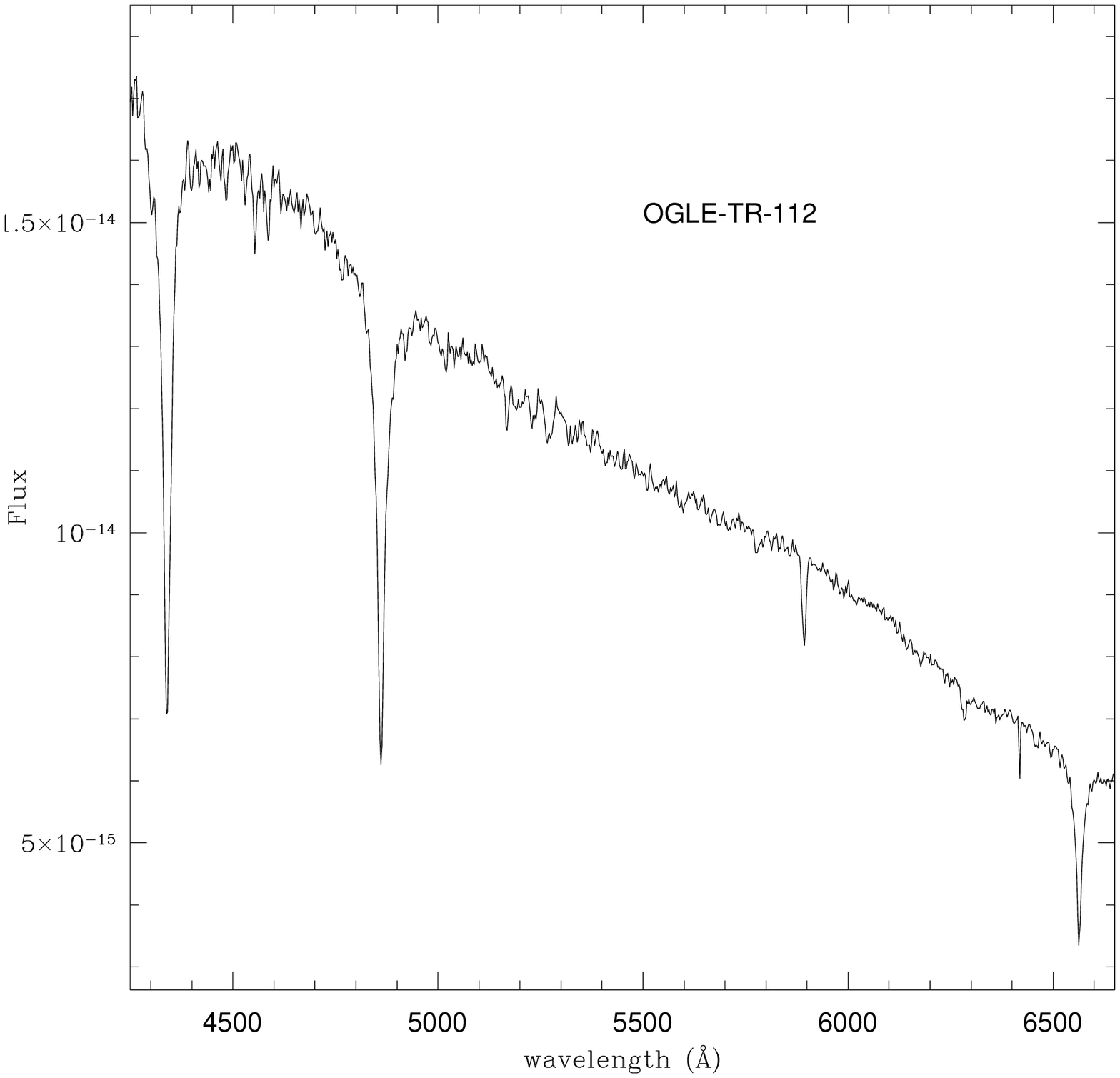,height=50mm}
\psfig{file=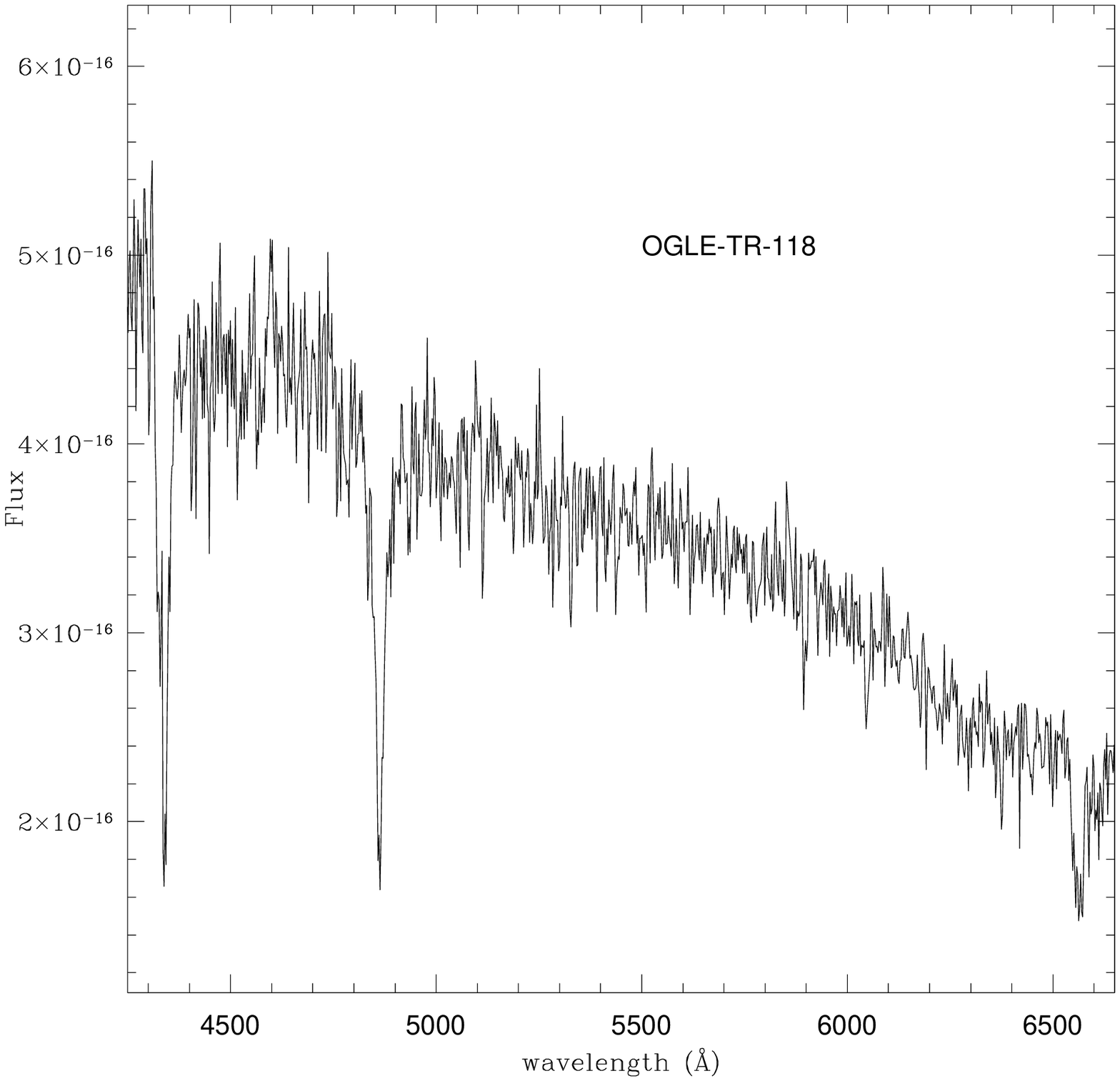,height=50mm}
\end{center}
\caption{Spectra of selected OGLE stars with transit candidates.}
\label{fig:spectra}
\end{figure*}

\begin{figure*}[h!]
\centering
\includegraphics[width=15cm]{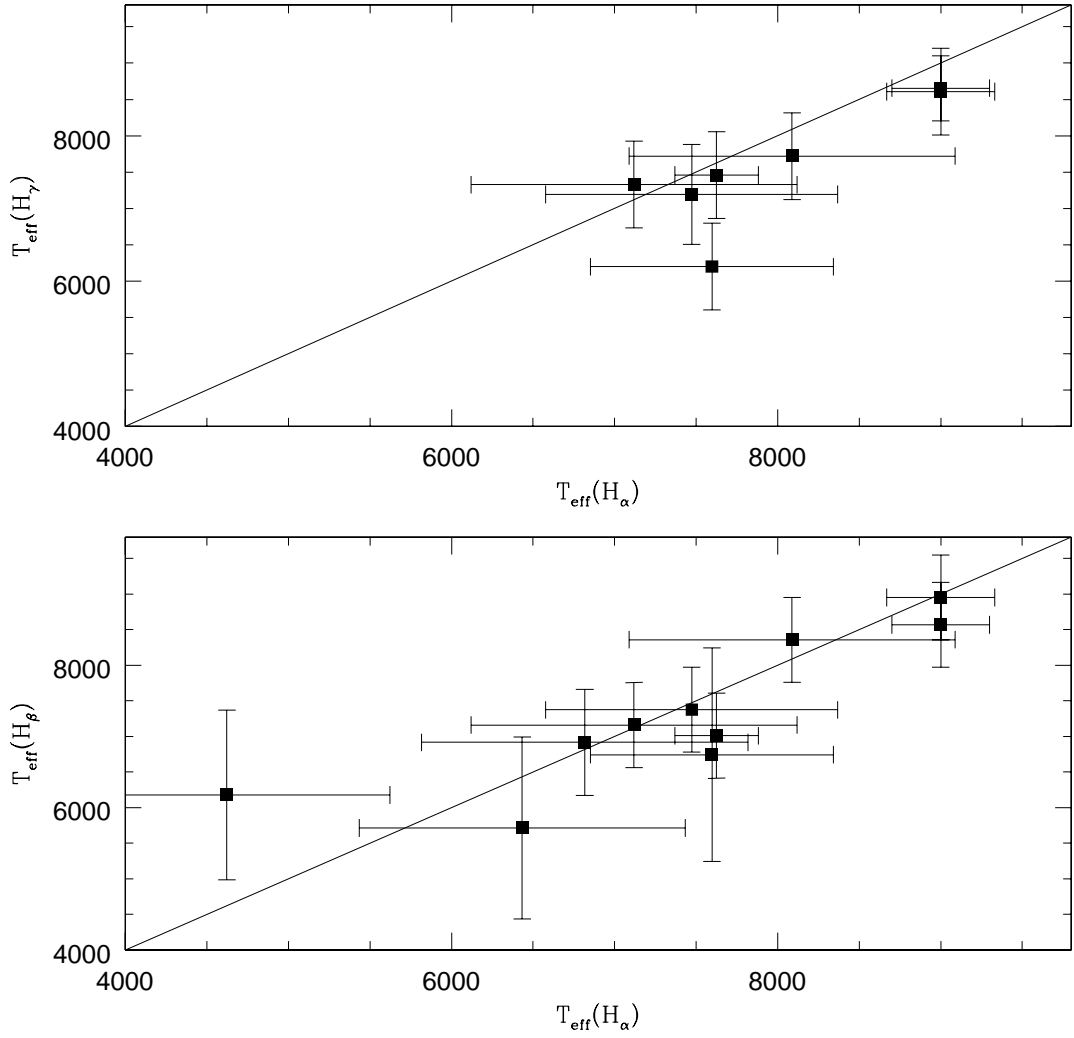}
\caption[]{Comparison of effective temperatures derived from the equivalent
widths of the Balmer lines.}
\label{fig:teffbalmer}
\end{figure*}

From the images of the spectra we have subtracted a combined bias, trimmed and divided by a
 combined 
normalised flat-field obtained from high S/N quartz lamps. The spectra were extracted using IRAF
tasks within the TWODSPEC package, with their background properly removed. The wavelength 
calibration was
done with the HeNeAr lamps, using 14 to 20 lines. The cosmic rays were excised, and the spectra
 were flux
calibrated using the spectrophotometric standards. However the flux calibration is unreliable
 because of the 
presence of thin cirrus during the nights. The final spectra are plotted in 
Figure~\ref{fig:spectra}. Although a qualitative  spectral classification can be made by comparing
these spectra with a series of templates (e.g. Dreizler et al. 2002), we prefer to measure the
effective temperatures using the three most intense lines from the Balmer series. 
 The results are summarised in Table~\ref{table:teffbalmer},
and Figure~\ref{fig:teffbalmer} shows that the independent measures from different Balmer lines is
in excellent agreement. The weighted mean of the three effective temperatures is given in Column 8
of Table~\ref{table:teffbalmer}, while the measured radial velocities are given in Column 9.

\section{Determination of stellar parameters}
%--------------------------------------------

\begin{figure*}[h!]
\centering
\includegraphics[width=15cm,angle=-90]{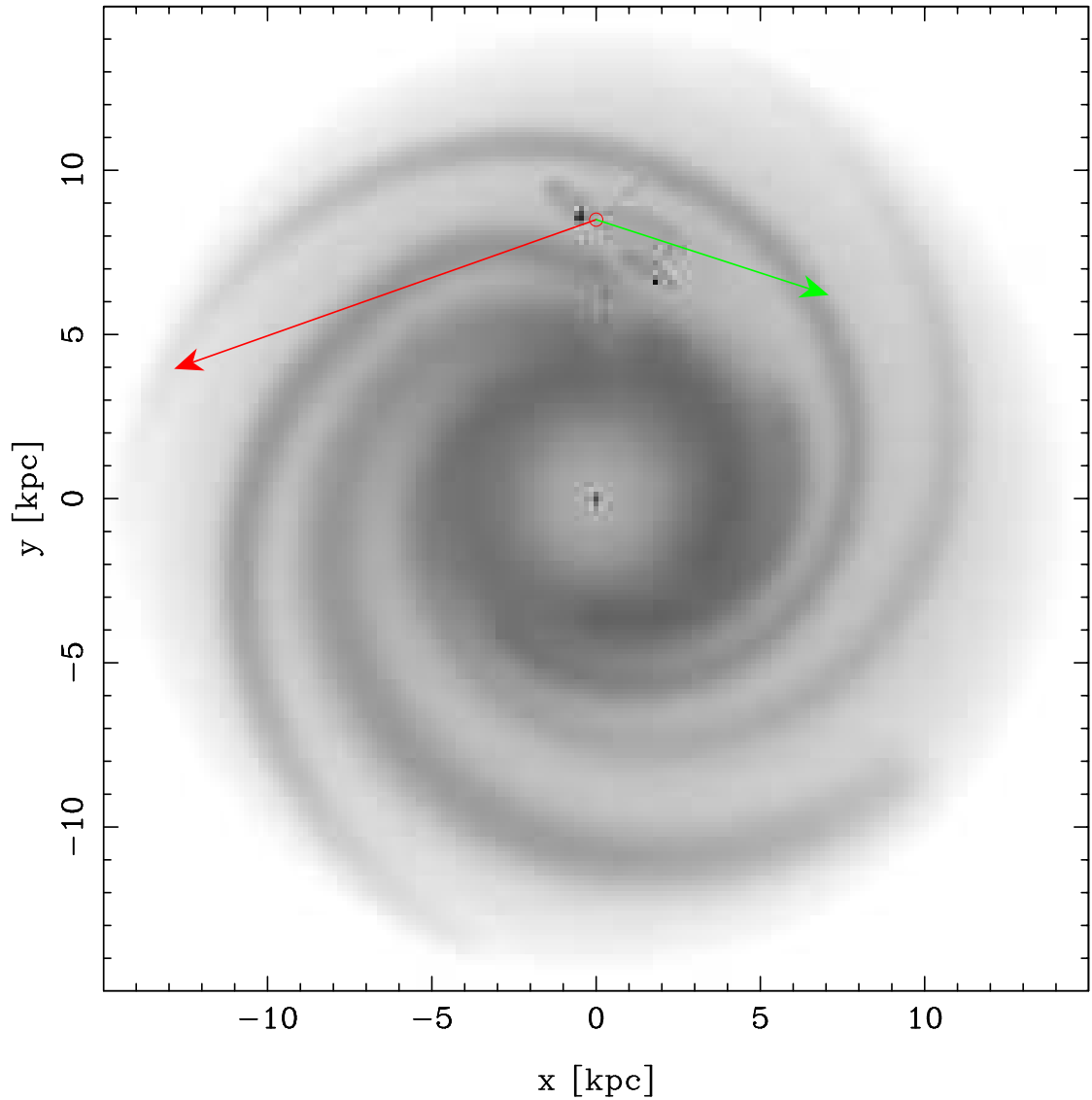}
\caption[]{Location of the Carina fields (leftwards arrow) and the field proposed for
the Kepler mission 
(rightwards arrow)
within the spiral arms of the Galaxy, as traced by the distribution of free electrons 
(Cordes \& Lazio 2003).  
The Carina line of 
sight clearly crosses the Carina-Sagittarius arm in at least two places, at about 2 and 7 kpc
 from the Sun. 
In comparison, the Kepler field crosses the Perseus arm at a much larger distance of 6 kpc.}
\label{fig:carina}
\end{figure*}

The determination of the properties of the stars which have transits, based solely on
optical and $K$-band photometry and on low-resolution spectroscopy, is complicated by the
lack of information on the extinction along their lines of sight. Even though the Carina
fields (Table~\ref{table:fields}) correspond to regions of the Galactic disc where the reddening 
and extinction are relatively low, locally, they cross the Carina-Sagittarius spiral arm at least twice,
as shown on Fig.~\ref{fig:carina} where the arms were traced with the distribution of free
 electrons
as determined by dispersion measures on pulsars (Taylor and Cordes 1993, Cordes and Lazzio 2003).
Extinction is extremely patchy in these regions of the Carina arm (Wramdemark 1980), which are
populated by many star-forming
complexes and cavities (Georgelin et al. 2000), leading to a wide range not only of
extinction values but also of the reddening parameter \rv, which in these areas ranges
from 3.0 to 5.0, with an average of about 4.0 (Patriarchi, Morbidelli \& Perinotto 2003), much
larger than the anomalous ones seen in the bulge (Udalski 2003).

\begin{figure*}[h!]
\centering
\includegraphics[width=12cm,angle=-90]{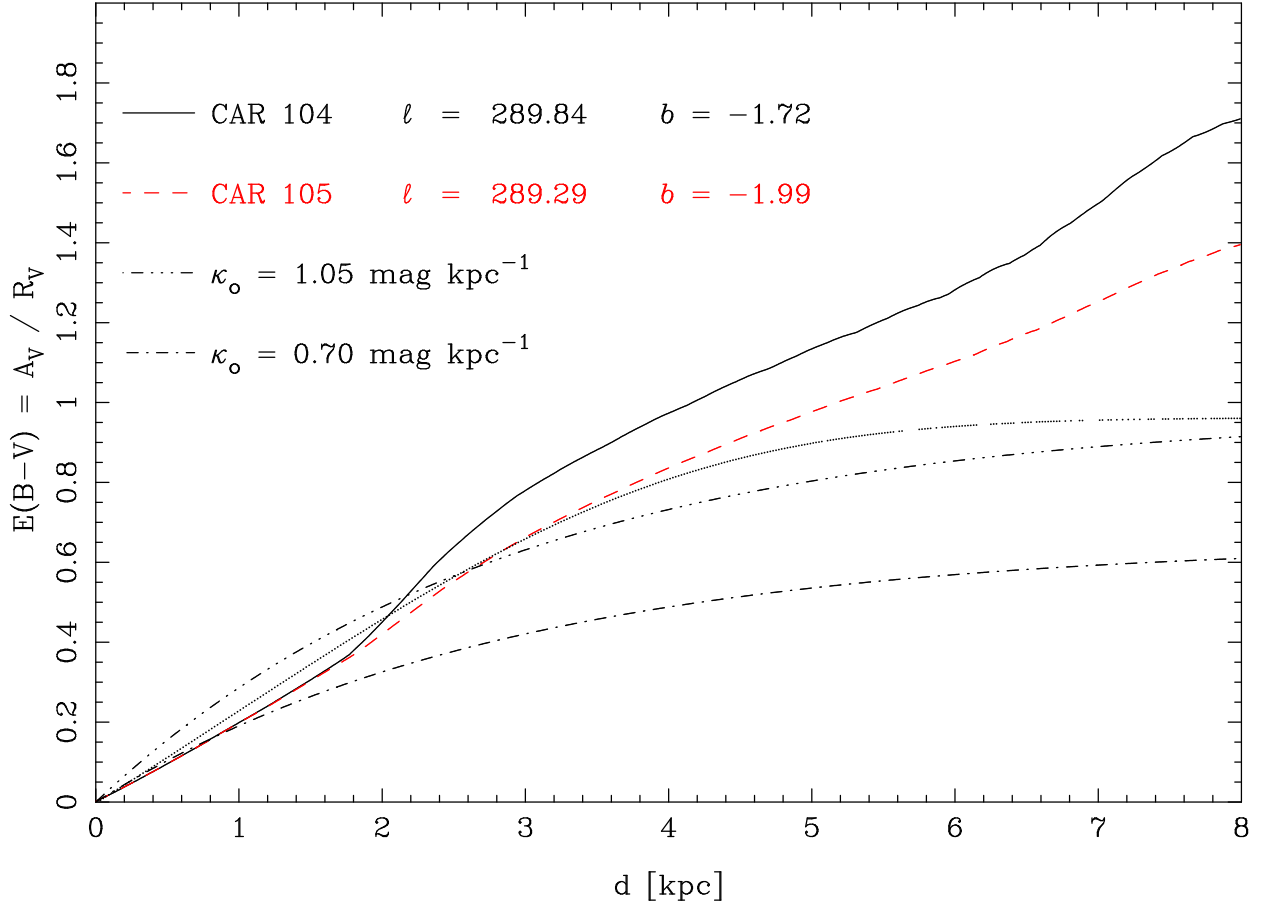}
\caption[]{Predicted colour excess E(B-V) along the lines of sight to the fields
CAR~104 (full line) and CAR~105 (dashed line), according to the 3-D model by Drimmel et al. 
(2003) and to the  Besan\c{c}on Galaxy model in these directions (dotted line, 
Robin et al. 2003). 
The predicted increases at 2 kpc and 7 kpc in the former model correspond to the crossing of 
the Carina spiral arm  (see Fig.~\ref{fig:carina}). For reference, two simple exponential
models are also shown to indicate the likely range of reddening at a given distance. In
all cases $R_V=3.1$ is assumed.}
\label{fig:ebvd}
\end{figure*}

Since the target stars presumably lie, given their magnitudes and spectral types, on the 
first intersection with the Carina spiral arm, we 
estimate the reddening
along the line of sight with the 3-dimensional model by Drimmel et al. (2003) as a guide. 
Figure~\ref{fig:ebvd} shows
the colour excess \ebv~ as a function of distance (assuming \rv~= 3.1) predicted by this model
for the two fields considered in this study, along with the one assumed by the Besan\c{c}on 
Galaxy
model which we will use later on to analyse the field population (Section 6). This improves upon 
the estimation made by Mendez \& van Altena (1998), and appears to be the only practical guide
at low galactic latitudes. The Drimmel et al. (2003) 3-dimensional model predicts two
important features long these directions (see Fig.~~\ref{fig:ebvd}). First, the intersection
with the spiral arm is reflected by the rapid increase in the absorption at about 2 kpc, and then at about
6.5 kpc, fully consistent with the map shown on Fig.~\ref{fig:carina}. The Besan\c{c}on Galaxy model 
is, in comparison, less realistic as this feature is not present. Second, the predicted 
colour excess at 8 kpc appears extremely large, between 1.4 and 1.7 magnitudes in \ebv, while
the Besan\c{c}on Galaxy model reaches an asymptote well below 1 magnitude. The reason of this
discrepancy is unclear, but may perhaps be related to the fact that the 3-D model is calibrated on the 
data of DIRBE/COBE and ISSA/IRAS which may overestimate the reddening by factors up to 1.5
in regions where \av$ > 0.5 $ mag (Arce \& Goodman 1999). We also note that simple,
homogeneous exponential discs will produce an extinction of the form
\beq
A_{\mathrm{V}} \; = \; \left(\frac{ \kappa_o }{10 \, \sin |b|}\right) \, 
                       \left( 1 - \exp\left[ \, - 10 \, d \, \sin |b|\right] \right)
\eeq
where $d$ is measured in kpc, so that asymptotically when $b \to 0$ the extinction tends
to $\kappa_o$ magnitudes per kpc. The Besan\c{c}on Galaxy model appears to give a distance
dependent extinction similar to an exponential disc with $\kappa_o = 1.05$ (Fig.~\ref{fig:ebvd}).
To simulate and account for the effect of a possible increase of \rv~ as large as 4, we will also 
use  $\kappa_o = 0.70$.
Note that while the extinction predicted by this series of models at large distances is very
different, below 2 kpc the range becomes rather narrow, and the results will largely be
independent of the assumed extinction. At large distances, the unknown dispersion of \av~ at
a given distance will be a limiting factor.

The simultaneous measure of \rv~ and \ebv~ independently would require a much better absolute flux 
calibration of the low resolution spectra and $U$-band photometry or UV spectra,  and is clearly 
beyond the scope of this work. 
We will assume instead \rv$= 3.1$ and consider a variety of extinctions to account both for a possible
variation of \rv~ and the dispersion of extinction at a given distance.
To determine the
extinction of each star, the colour-colour diagrams (Fig.~\ref{fig:colcol}) cannot be used, because the
reddening vector happens to have the same direction as the locus of the main sequence stars, and
even some giants may also lie close to the sequence.

\begin{figure*}[h!]
\centering
\includegraphics[width=12cm,angle=-90]{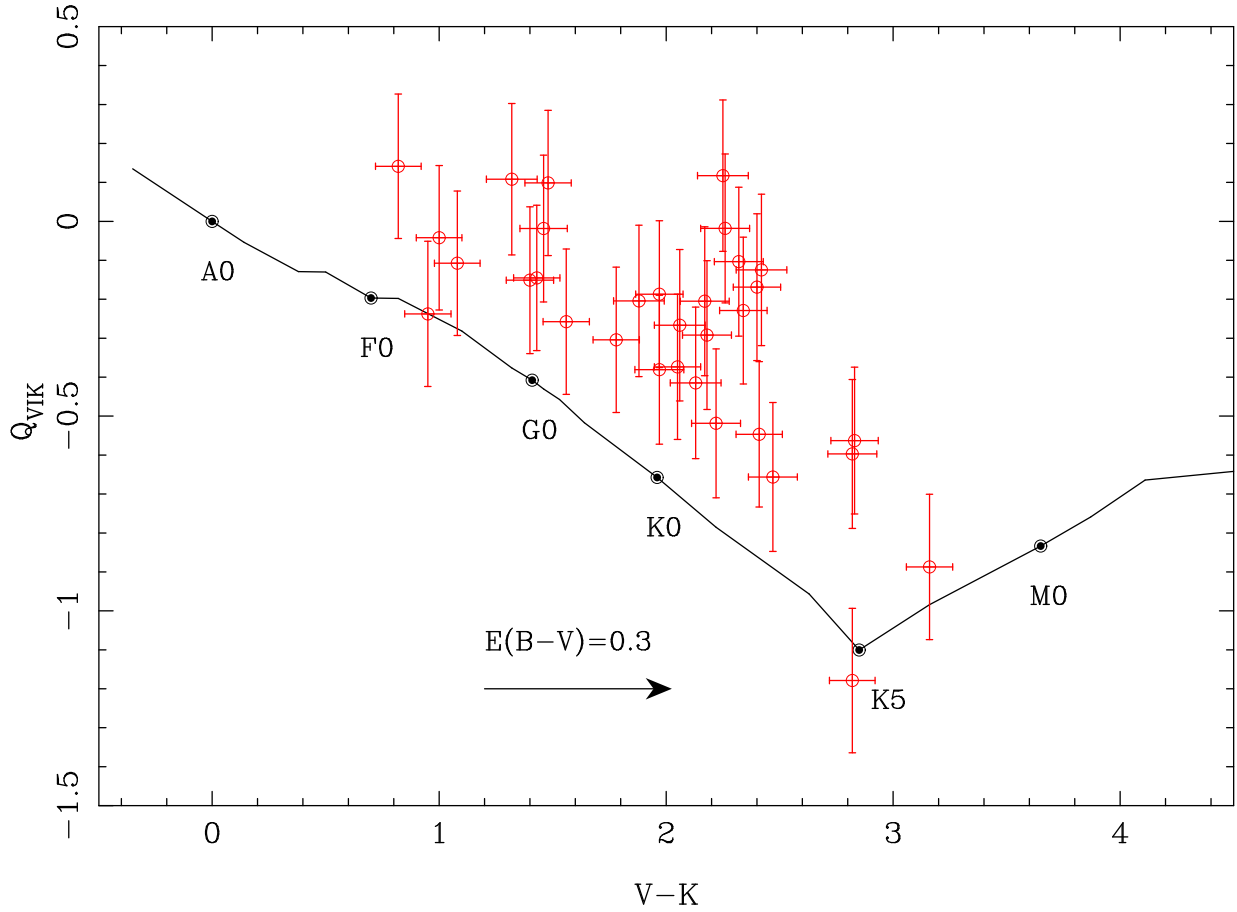}
\caption[]{Reddening-free index $Q_{\mathrm{VIK}}$ as a function of the observed $V-K$ colour.
 The solid line gives the theoretical locus of unreddened main sequence stars. Note the 
systematic horizontal offset produced by reddening, whose average colour excess E(B-V) of about 0.3 
magnitudes is indicated by the horizontal arrow.}
\label{fig:qvik}
\end{figure*}

One possibility would be to compute reddening-independent indices based on the standard
dependence of extinction with wavelength. For instance,
\beq
 Q_{\mathrm{VIK}} \;  = \;  (V-I) \, -  \, \left[ \frac{E(V-I)}{E(I-K)} \right] \, (I-K)
\eeq
where
\begin{eqnarray}
E(V-I) \; & = & \; A_{\mathrm{V}} \, \left( 1.0 - \frac{A_{\mathrm{I}}}{A_{\mathrm{V}}} \right) \\
E(I-K) \; & = & \; A_{\mathrm{V}} \, \left( \frac{A_{\mathrm{I}}}{A_{\mathrm{V}}}  - 
                           \frac{A_{\mathrm{K}}}{A_{\mathrm{V}}} \right) 
\end{eqnarray}
and we adopt $A_{\mathrm{I}}/A_{\mathrm{V}} = 0.479$ and $A_{\mathrm{K}}/A_{\mathrm{V}} = 0.113$ 
(Cardelli et al. 1989). 
The resulting diagram is shown on Fig.~\ref{fig:qvik} where the horizontal shift produced
by extinction is clearly visible and amounts, in average, to about \ebv $=0.3$. If all the stars
were on the main-sequence, the reddening would be measured through the horizontal shift required
to move the observed position to the main sequence locus. Unfortunately we have no guarantee
that all these stars are on the main sequence of solar metallicity, as some may be more metal
rich and some giants could well be present, especially in the lower half of the diagram. Clearly
another technique has to be used to measure unambiguosly the reddening.

In this paper we introduce a new method based on the well-known surface brightness (SB) relation which 
has been recently recalibrated using exclusively direct interferometric measures of angular 
diameters (Kervella et al. 2004). The surface brightness $S_{\lambda}$ in a given passband is
just
\beq
S_{\lambda} \; = \; a \, - \, 0.1 m_{{\lambda}_o} \, - \, 0.5 \log \, \theta_{\mathrm{LD}}
\eeq
where $m_{{\lambda}_o}$ is the intrinsic, dereddened apparent magnitude in that passband and
$\theta_{\mathrm{LD}}$ its limb-darkened angular diameter. Since the surface brightness is tighly correlated
to colour, the equation can be inverted to yield the diameter as a function of colour.
Our method has thus three distinct steps. First, a distance is assumed, and hence a reddening (based
on the model adopted for extinction). Magnitudes and colours are dereddened. The resulting optical and
IR magnitudes are used to infer the limb darkened angular diameter of the star, 
via the relations
\begin{eqnarray}
\log \, \left( \frac{\theta_{\mathrm{LD}}}{\mathrm{mas}} \right)  & = &  0.5170 \, + \, 0.0755 (V-K)_o \, - \, 0.2 K_o \\ 
\log \, \left( \frac{\theta_{\mathrm{LD}}}{\mathrm{mas}} \right)  & = &  0.5149 \, + \, 0.1805 (I-K)_o \, - \, 0.2 K_o 
\end{eqnarray}
where the  rms dispersions are less than 1\% and 3.7\% respectively (Kervella et al. 2004). 
Our final estimate is the weigthed mean of these two estimates. 

The second step estimates the effective temperature through the relation 
\begin{eqnarray}
\lefteqn{ \log T_{\mathrm{eff}} \;  =  \;  4.1788 }   \\ \nonumber
         & &  - \, \left[ 1.1806 \log \theta_{\mathrm{LD}} \, + \, 0.2361 K_o 
\, - \, 0.5695 \right]^{1/2}
\end{eqnarray}
with a dispersion smaller than 0.6\%, which corresponds to a systematic error of 40 K for a G2 V star 
(Kervella et al. 2004). The combination of the first two steps yields an estimate of the effective
temperature which depends on the assumed distance. It also yields directly the intrinsic stellar radius as
\beq
\left(\frac{R_*}{R_\odot}\right) \; = \; 4.434 \, 10^{10} \; \left( \frac{d}{\mathrm{kpc}} \right) 
 \; \tan \left( \frac{ \theta_{\mathrm{LD}} }{2} \right)
\eeq

The third and last step makes use of the properties of main-sequence stars : the strong correlation
between luminosity and mass ($L \sim M^{\beta}$), and between mass and radius ($M \sim R^{\alpha}$),
and hence between effective temperature and radius of the form
\beq
\left( \frac{ T_{\mathrm{eff}} }{T_{\mathrm{eff}\odot}} \right) \; = \; 
\left( \frac{ R_* }{R_{\odot}} \right)^{\left( \alpha \beta - 2 \right)/4} \; \approx \; 
\left( \frac{ R_* }{R_{\odot}} \right)^{0.64}
\label{eq:ms}
\eeq
using the standard values of $\alpha=1/0.8$, $\beta \approx 3.6$ and $T_{\mathrm{eff}\odot} = 5770 $ K. 
The resulting dependence of this new estimate of the effective temperature with the distance will
clearly be different from the one resulting from the surface brightness relation. If a distance
exists such that both estimates agree, we have a full solution with all the parameters of a
main sequence star that agrees with the optical and IR photometric data. If there is no solution,
then either the star is a giant, for which the surface brightness relations that we have
used do not apply, or the assumed reddening at that distance is incorrect.

\begin{figure*}[h!]
\centering
\includegraphics[height=15cm,angle=-90]{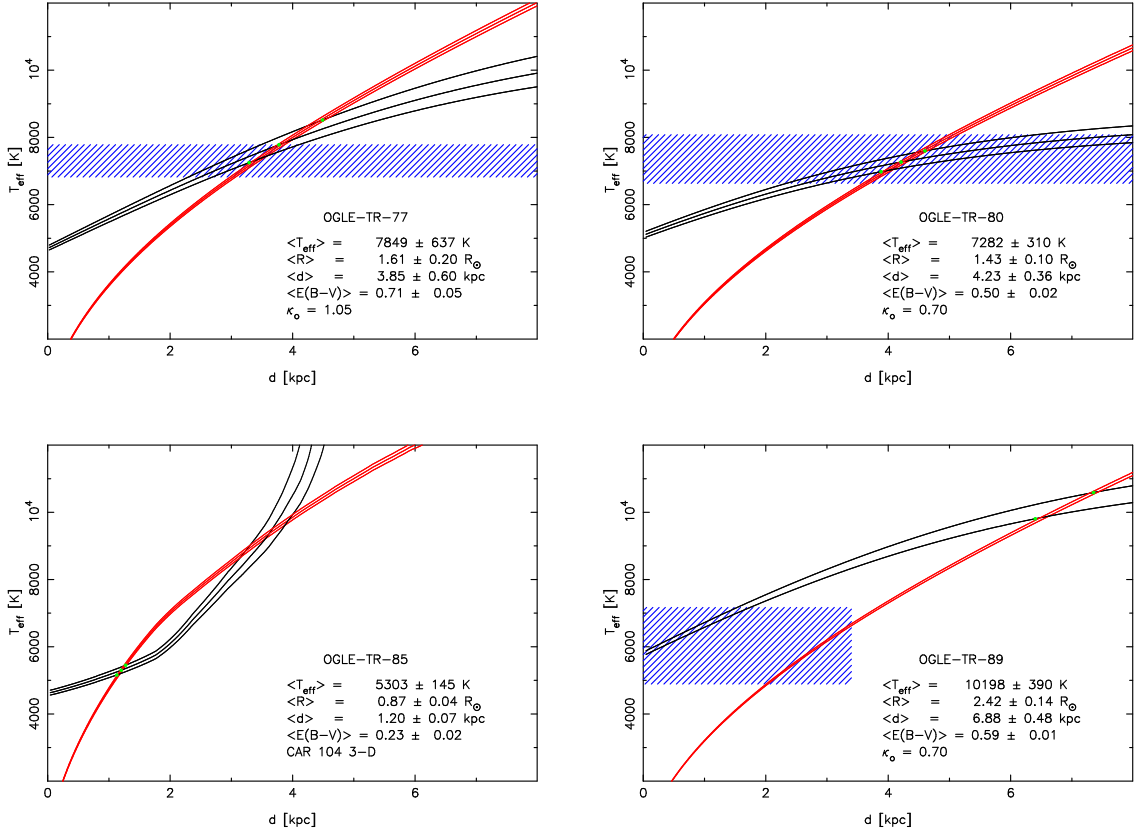}
\caption[]{Effective temperature versus distance for selected candidates (\ogle77, \ogle80,
 \ogle85, \ogle89). The red lines
show the radius-temperature constrain for main sequence stars, weakly dependent on 
photometrical errors, while the black lines give the
range of temperatures inferred from the surface brightness relation when the errors
in the $V-K$ colours and $V$ magnitudes are considered. If present, the blue hatched region
give the range of effective temperature allowed by the main Balmer lines. 
A colour figure is available in the electronic version of the paper.}
\label{fig:sbiter1}
\end{figure*}

The procedure is illustrated in Fig.~\ref{fig:sbiter1} for 4 candidates from the OGLE list. In the
case of \ogle{77}, the three black lines give the effective temperature as a function of distance 
(assuming $\kappa_o = 1.05$) for the mean and the two extreme values of the optical and K-band
magnitudes. The relatively large separation between these three estimates is due to the large
uncertainties (0.10--0.15 mag) in the OGLE optical magnitudes. The red lines give the
constraint from the main sequence (Eq.~\ref{eq:ms}) which appears to be far less sensitive to the
photometrical errors. The intersections of the lines yield a unique and accurate  solution for the
effective temperature, the stellar radius, the distance and the reddening. 

\begin{figure*}[h!]
\centering
\includegraphics[height=15cm,angle=-90]{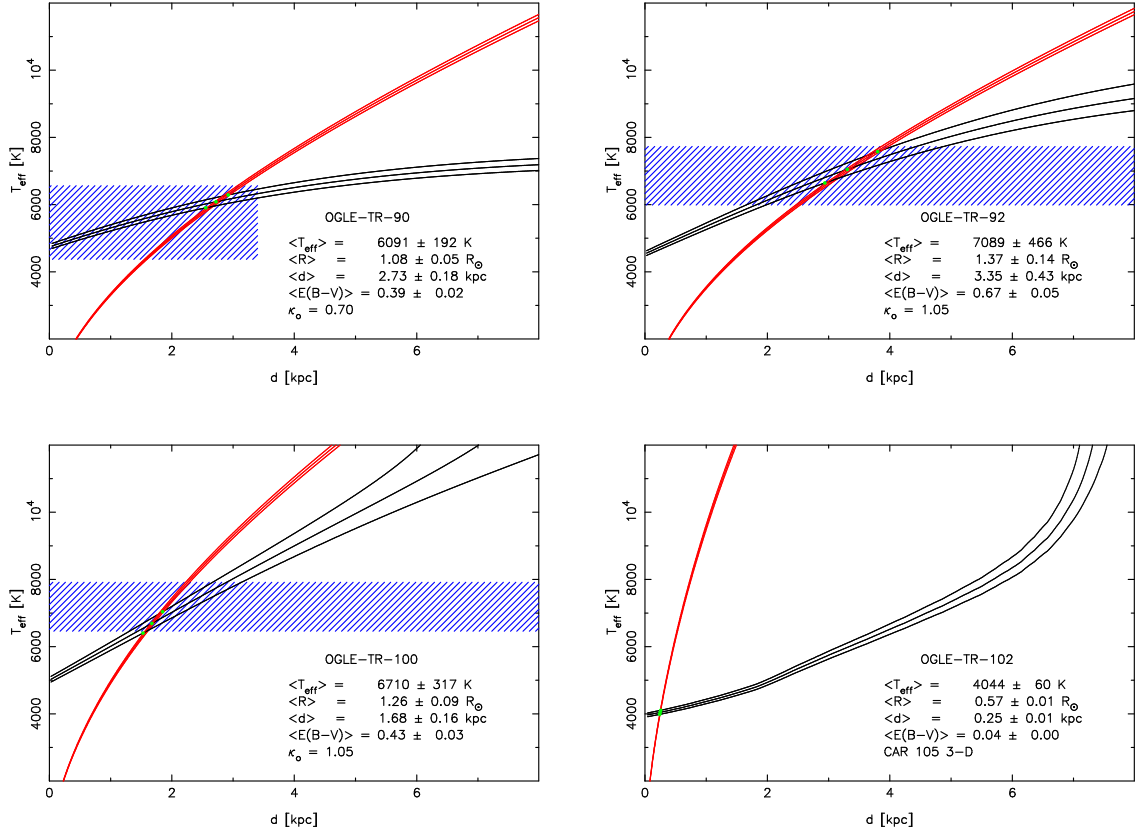}
\caption[]{Same as Fig. \ref{fig:sbiter1} for \ogle{90}, \ogle{92}, \ogle{100} and \ogle{102}. 
A colour figure is available in the electronic version of the paper.}
\label{fig:sbiter2}
\end{figure*}

\begin{figure*}[h!]
\centering
\includegraphics[height=15cm,angle=-90]{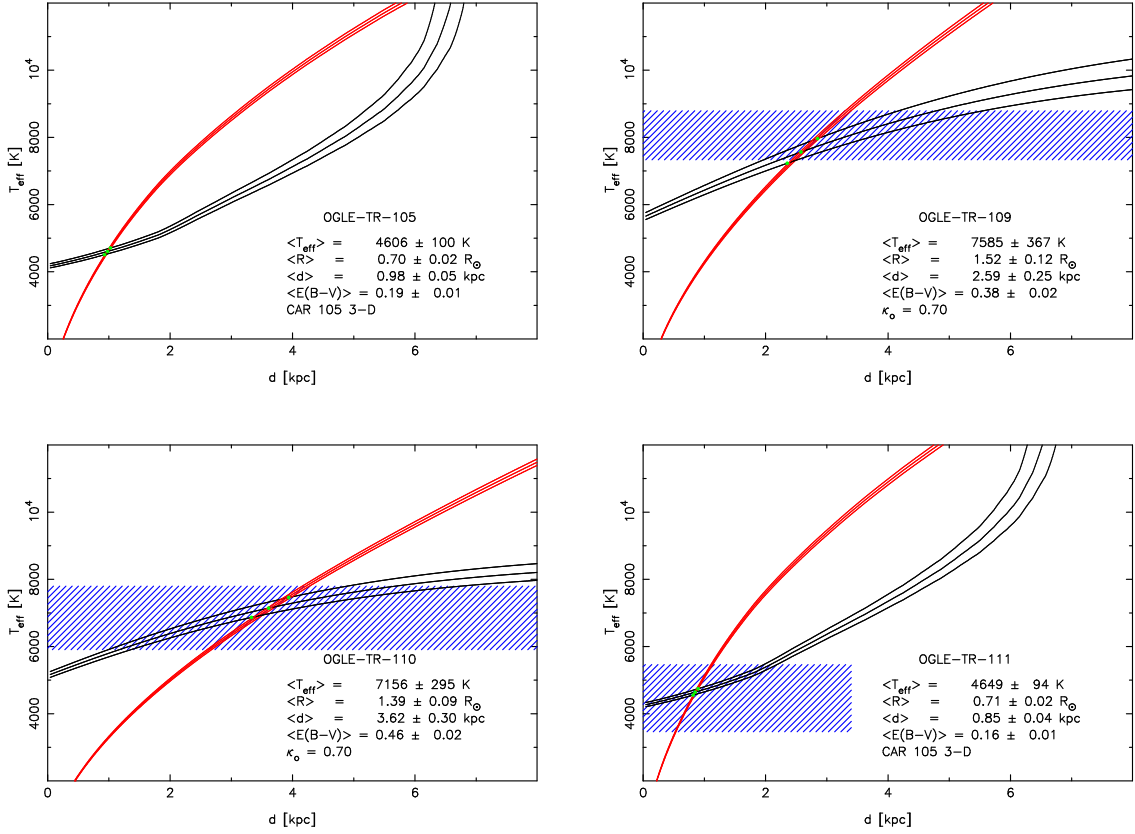}
\caption[]{Same as Fig. \ref{fig:sbiter1} for \ogle{105}, \ogle{109}, \ogle{110} and \ogle{111}. 
A colour figure is available in the electronic version of the paper.}
\label{fig:sbiter3}
\end{figure*}

\begin{figure*}[h!]
\centering
\includegraphics[height=15cm,angle=-90]{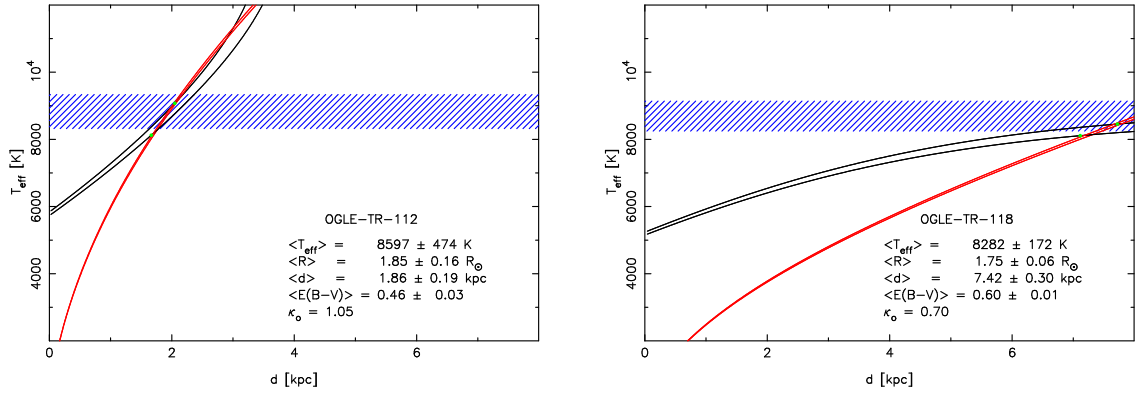}
\caption[]{Same as Fig. \ref{fig:sbiter1} for \ogle{112} and \ogle{118}. 
A colour figure is available in the electronic version of the paper.}
\label{fig:sbiter4}
\end{figure*}

A fully independent estimate comes from the measures of the effective temperatures based on the
low-resolution spectra (Table~\ref{table:teffbalmer}). When available, these measures are indicated
on Figs.~\ref{fig:sbiter1} to \ref{fig:sbiter4} as a hatched band, independent of distance. 
As shown on Fig.~\ref{fig:sbiter1} the agreement for \ogle{77} and \ogle{80} is impressive.

The method is not devoid of problems, however. As  Fig.~\ref{fig:sbiter1} illustrates, in the case
of \ogle{85} two solutions are possible, and there is no way to select the correct one on the
basis of that information only. However, the value of the reddening-free index $Q_{\mathrm{VIK}}$
indicates that \ogle{85} must have a spectral class between G2 and K0 (Fig.~\ref{fig:qvik}) and
hence the short distance solution is preferred. The steep increase in the SB-derived temperatures
in this case was produced by the steep increase in extinction predicted by the 3-D extinction
model by Drimmel et al. (2003). Other models would not give the second solution, as at that
distance the expected scatter in reddening is very small, and so, again, the short distance solution
is preferred. 

A second problem arises when the temperatures derived from the SB relation do not agree with the
spectroscopic ones. This is the case for \ogle{89} (Fig.~\ref{fig:sbiter1}) where the SB-inferred
temperature gives a temperature of some 10,000~K while the spectrum indicates 6,000~K. Again the
reddening-free index $Q_{\mathrm{VIK}}$ (see Fig.~\ref{fig:qvik}) indicates that the position 
of \ogle{89} is consistent with the locus of A-type stars, pointing either to a problem in the spectrum 
or else that it is a giant star, in which case the SB-inferred temperature is of course incorrect.
The latter explanation is likely, as the SB method points to a large distance (6.9 kpc) which
is very unlikely for this bright star.

While at distances below 2~kpc the inferred measures are independent of the reddening assumed
at these distances (Fig.~\ref{fig:ebvd}), at larger distances there is a rather strong dependence
on the assumed extinction. For instance, in the case of \ogle{89} or \ogle{118}, an extinction larger than the
one predicted by the $\kappa_0=0.7$ exponential model would yield either a very large distance,
which is unlikely given their magnitudes, or no solution at all, pointing to  supergiant stars. 

\begin{figure*}[h!]
\centering
\includegraphics[height=12cm,angle=-90]{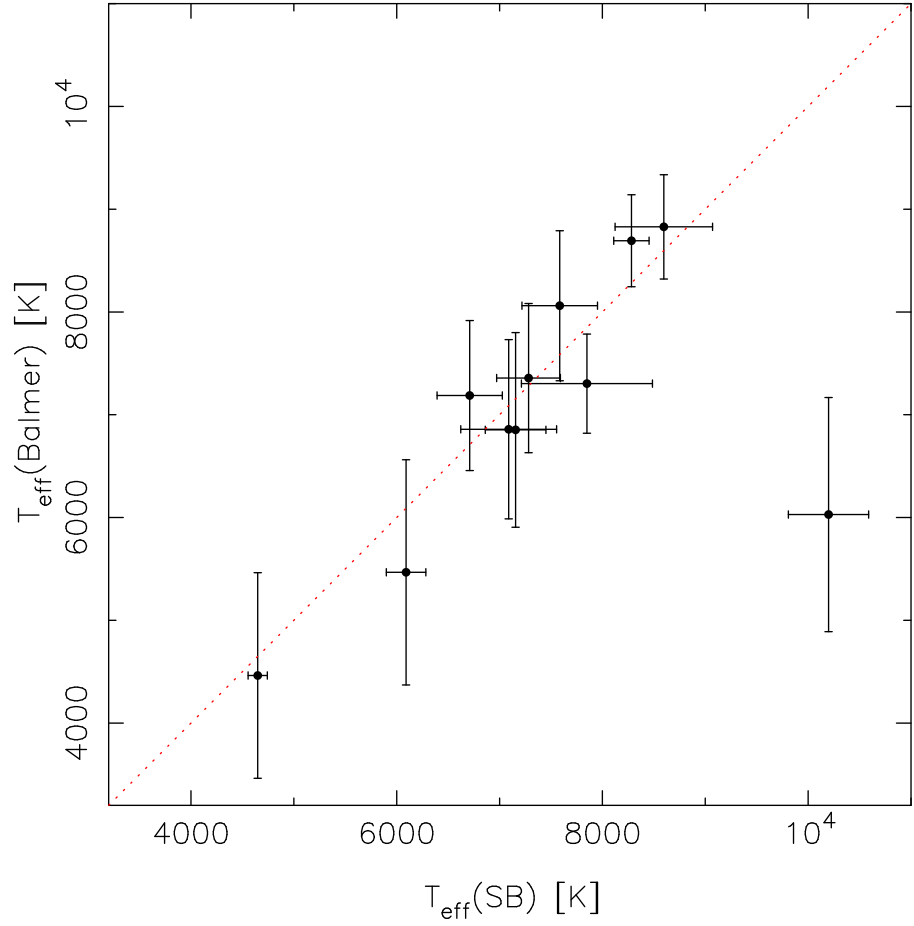}
\caption[]{Comparison of effective temperatures derived from 
the Balmer lines and from the surface brightness method. The only outlier
is \ogle{89}, which is very likely a giant star.}
\label{fig:teff_sb_balmer}
\end{figure*}

In general, the agreement between the SB-inferred temperatures and the spectroscopic ones is very good,
as illustrated on Figure~\ref{fig:teff_sb_balmer}. The only outlier is, in fact, \ogle{89} which is
likely to be a giant star, as discussed above. Likewise, the large distance inferred for 
\ogle{118} (see Fig.~\ref{fig:sbiter4}) makes it likely to be a giant star. 

Some other stars deserve comments. \ogle{102} (Fig.~\ref{fig:sbiter2}) appears to have a solution
with a very short distance (0.25 kpc), as appropriate for such a low temperature star inferred
from the SB method. This temperature is fully consistent with its small  $Q_{\mathrm{VIK}}$ index
which makes it a late main-sequence K star (Fig.~\ref{fig:qvik}). Yet its
position on the colour-magnitude (Fig.~\ref{fig:KVKcmd}) shows that it could also be a giant
star, at of course a much larger distance. The Galaxy model that we will use in \S 6 shows that
such a small distance is not expected in a sample of stars of this size, providing yet another
hint that this star is a giant. 

In summary, \ogle{83}, \ogle{98}, \ogle{116} and \ogle{119} do not yield any solution and are
therefore very likely to be giant stars. As their main-sequence solutions yield large distances, 
\ogle{89} and \ogle{118} are also likely to be giants. \ogle{102} is also likely to be a giant, even
though a reasonable value for its parameters, consistent with all the data, is found assuming
that it is a main-sequence star. Table~\ref{table:sbfinal} summarises the stellar properties
inferred from our method.

Further refinements would require intermediate resolution spectroscopy. For instance, the measure of 
the KP (metallicity-dependent) and HP2 (pseudo-EW of the Balmer H$_{\delta}$ line) indices would 
allow 
the colour excess to be measured within 0.03 mag (Bonifacio, Caffau \& Molaro 2000)
independently of any assumption on distance (although \rv~ cannot be measured in this way).
Even though it is less
costly in telescope time than high resolution spectroscopy, it remains unpractical for a large
systematic follow-up survey. 

\section{Constraints on the companions}
%--------------------------------------
The main objective of our observations is to better characterise the properties of the host
stars in order to infer the radii of their companions. They are given straightforwardly  
through the standard relation (e.g. Seager \& Mall\'en-Orlenas 2003) 
\beq
\left( \frac{R_{c}}{R_{*}} \right) \; = \; \left( 10^{\Delta I / 2.5} - 1 \right)^{1/2} 
\label{eq:radius}
\eeq
where $\Delta I$ is the depth of the transit in magnitudes in the $I$ band, provided by OGLE
and given in Table~\ref{table:sbfinal}. 

Clearly the vast majority of companions are low-mass objects but are not small enough to be
exoplanets. If we adopt as an example the first extrasolar planet discovered with the transit
technique, HD 209458b (Charbonneau et al. 2000), which is a close giant 
planet\footnote{The radius of Jupiter is 0.103 $R_{\odot}$.} with a radius
of 1.35 $\pm$ 0.16 $R_{\mathrm{Jup}} \sim$ 0.14 $R_{\odot}$, only three companions could
be exoplanets :  \ogle{105}, \ogle{109} and \ogle{111}.

\ogle{102} is somewhat dubious because the host star is probably a giant star, as discussed above,
in which case the stellar radius has been seriously underestimated. In this case the radius
of the component has also been underestimated and the odds are that the transiting object is
not an exoplanet.

\ogle{105} and \ogle{111} are the best candidates, as the stellar radiii are measured to within
3\% and they are clearly main sequence stars with mid-K spectral types. \ogle{105} only had 3
transits measured (Udalski et al. 2002b) and therefore is a bit more dubious that \ogle{111} with
9 measured transits.

\ogle{109} would have the shortest period (below 1 day) and is a borderline object at $R_c$=0.13 
$R_{\odot}$,
along with \ogle{90} and \ogle{100}. They all share the property that their host stars are rather
hot, with effective temperatures around 7,000~K. Some of the MACHO transit candidates,
selected from a subset of 180,000 stars in the MACHO database of the Bulge, also 
appear around this type of stars (Drake \& Cook, 2004)

\begin{figure*}[h!]
\centering
\includegraphics[height=15cm,angle=-90.]{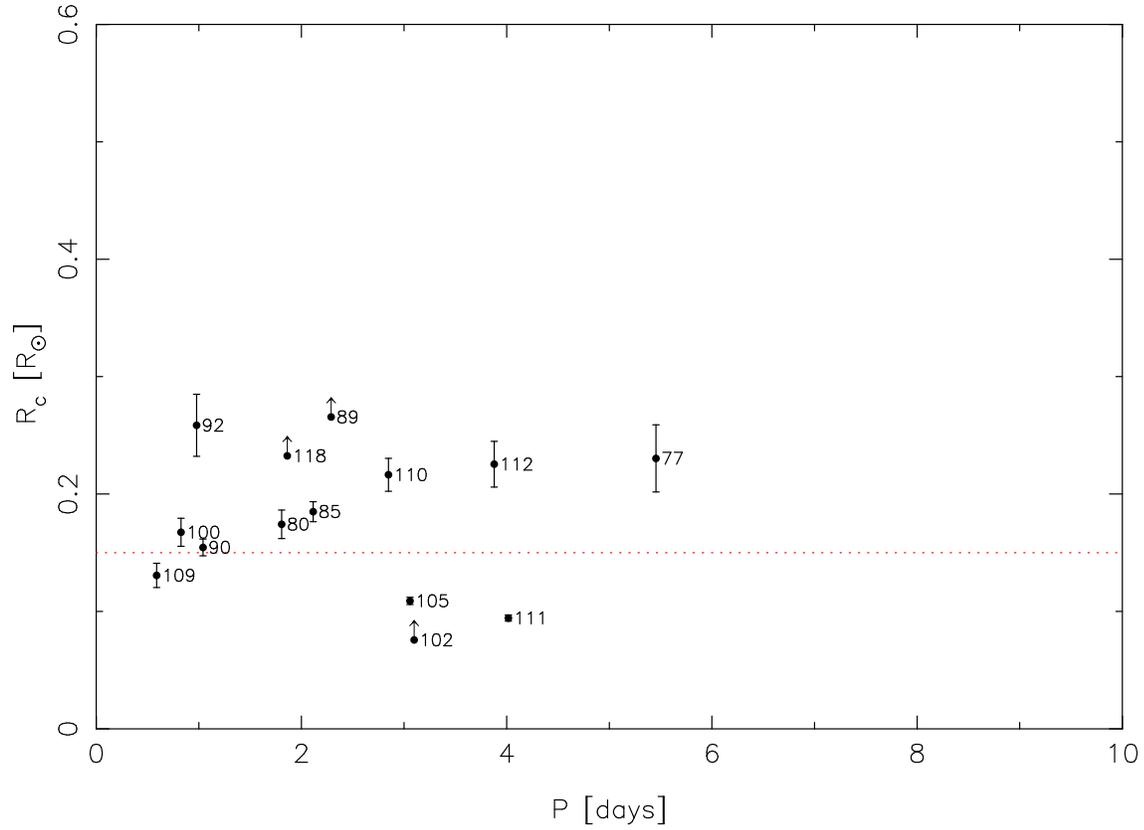}
\caption[]{Inferred radii for the companions $R_c$  
as a function of their orbital period. The dotted horizontal line at 0.15 $R_\odot$ marks
the adopted transition below which companions are likely to be exoplanets. The arrows
indicate the lower limits associated with likely host giant stars. Only
three transit candidates appear likely to have exoplanets : \ogle{105}, 
\ogle{109} and \ogle{111}.
}%
\label{fig:radius}
\end{figure*}

Figure~\ref{fig:radius} summarises the results by showing the inferred radii of the companions
as a function of their orbital period. The candidates selected here all fall within the
category of ``hot'' Jupiters as their periods are close to 3 days and appear to be quite
compact. It is beyond the scope of this work to discuss further the physical properties of this
subset of possible candidates.

It is interesting to compare the results of our analysis with the best candidates selected
by the OGLE team. They assumed that all the host stars had one solar mass exactly and inferred the
corresponding radii of the companions using Eq.~\ref{eq:radius}. Of their 10 best candidates with
$R_c/R_{\odot} < 0.14 $ that we observed, only two are confirmed : \ogle{109} and \ogle{111}, and
one is unclear. Overall, we exclude at least 70\% of their best candidates (with $R_c/R_{\odot} < 0.14 $)
and possibly 80\%. The are two reasons for this: in 30\% of the cases the star appears to be a giant,
while in the rest our revised radii become too large. In addition, one candidate which 
had $R_c/R_{\odot} = 0.16$
appears to have a smaller radius of 0.11 $R_{\odot}$ (\ogle{105}) and hence enters in the list
of possible exoplanets. This proves the effectiveness of our approach in pre-selecting the best
candidates for high-resolution spectroscopy.

\section{The stellar populations in the fields}
%----------------------------------------------

\begin{figure*}[h!]
\centering
\includegraphics[height=10cm,width=10cm]{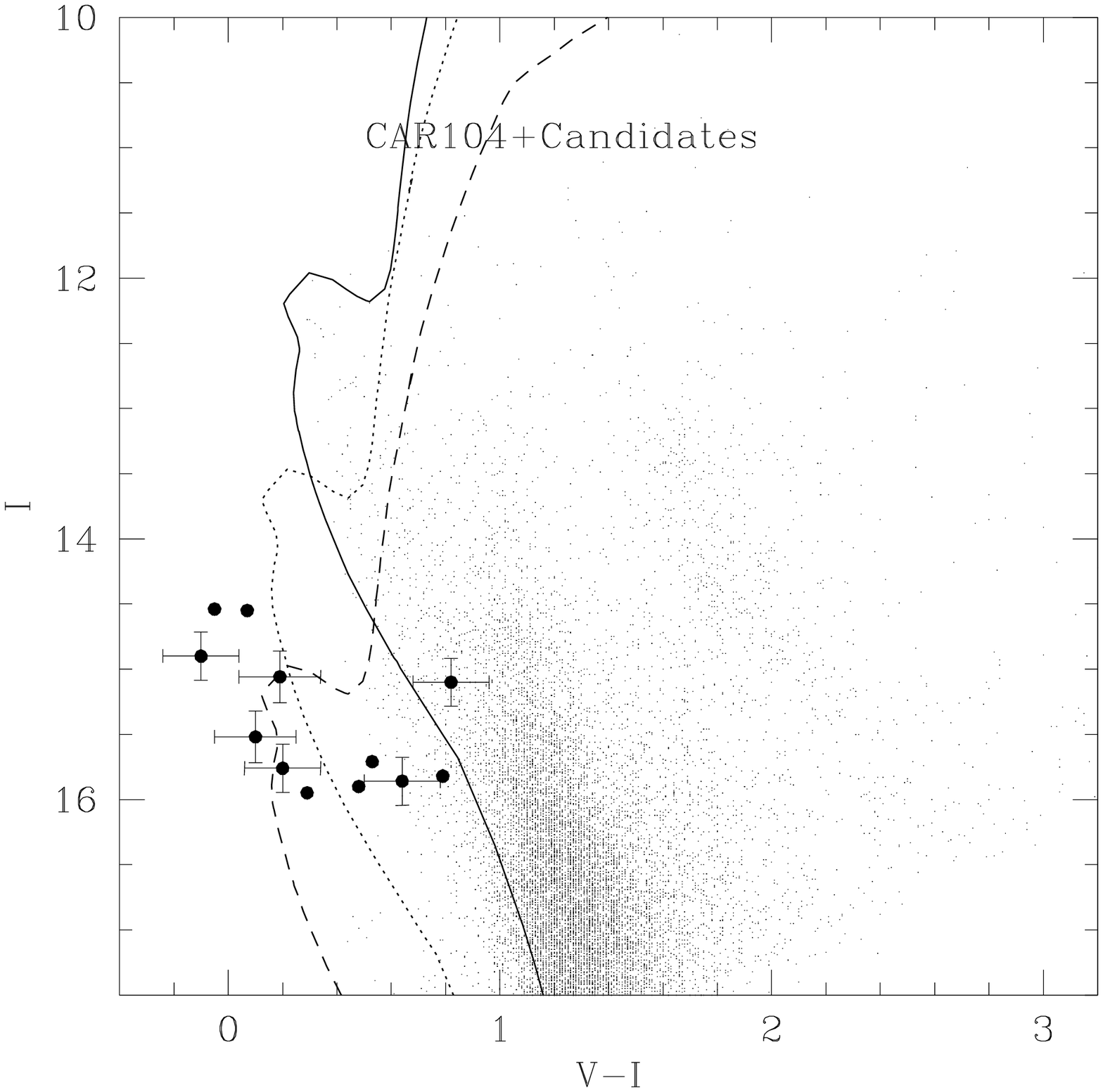}
\includegraphics[height=10cm,width=10cm]{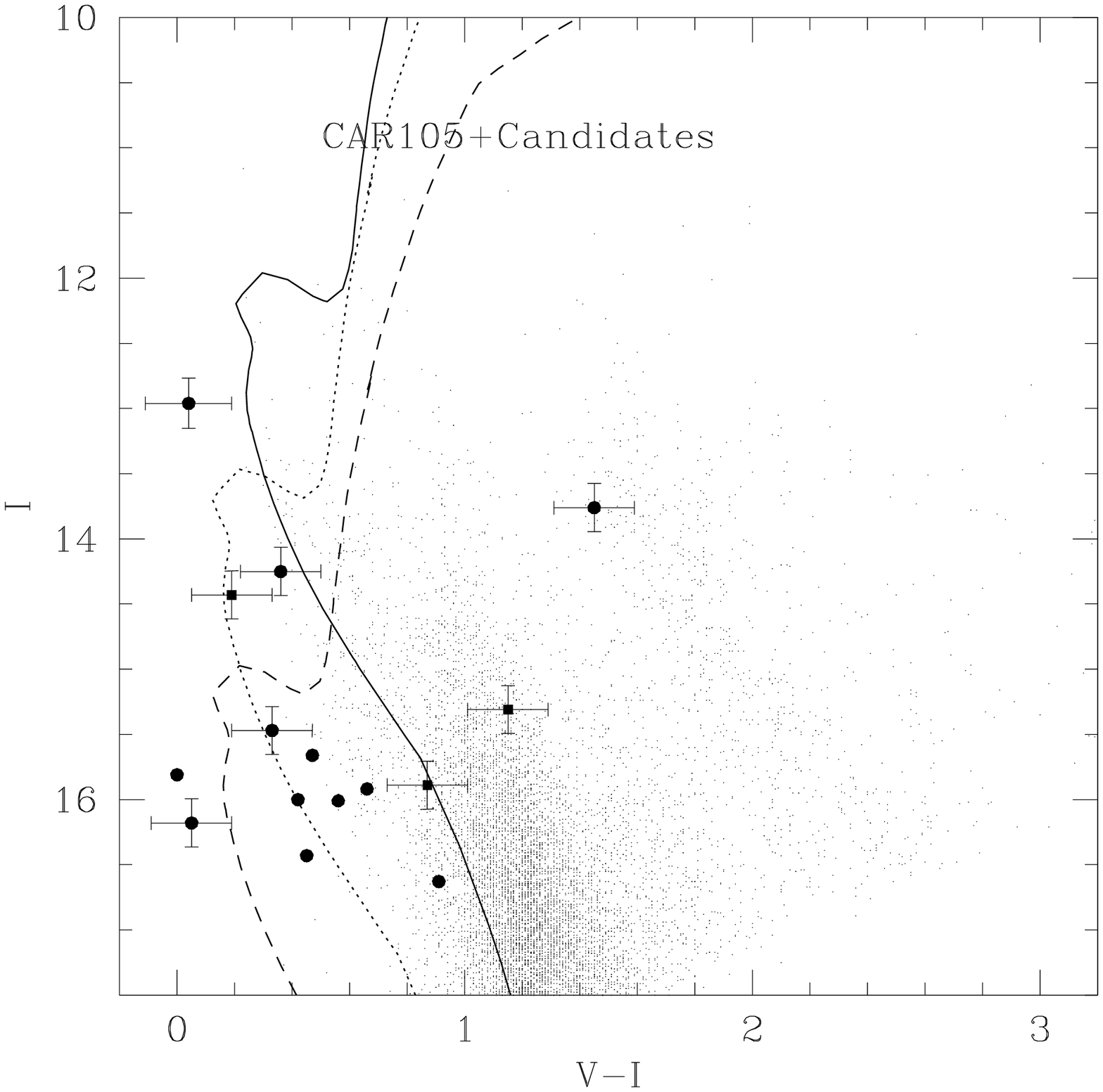}
\caption[]{$I$ vs $V-I$ colour-magnitude diagram for the CAR 104 and CAR 105 fields. The filled
 circles 
are the transit candidates. 
The lines show the Padova isochrones for solar age and 
metallicity (corrected for reddening), at
distances of 1 kpc (solid line), 2 kpc (dotted line) and 4 kpc (short dashed line). Note that, 
unlike the field stars, stars with transits have been corrected for reddening and 
that the most likely transit candidates to
host exoplanets are marked with full squares.}
\label{fig:IVIcmd}
\end{figure*}

\begin{figure*}[h!]
\centering
\includegraphics[height=10cm,width=10cm]{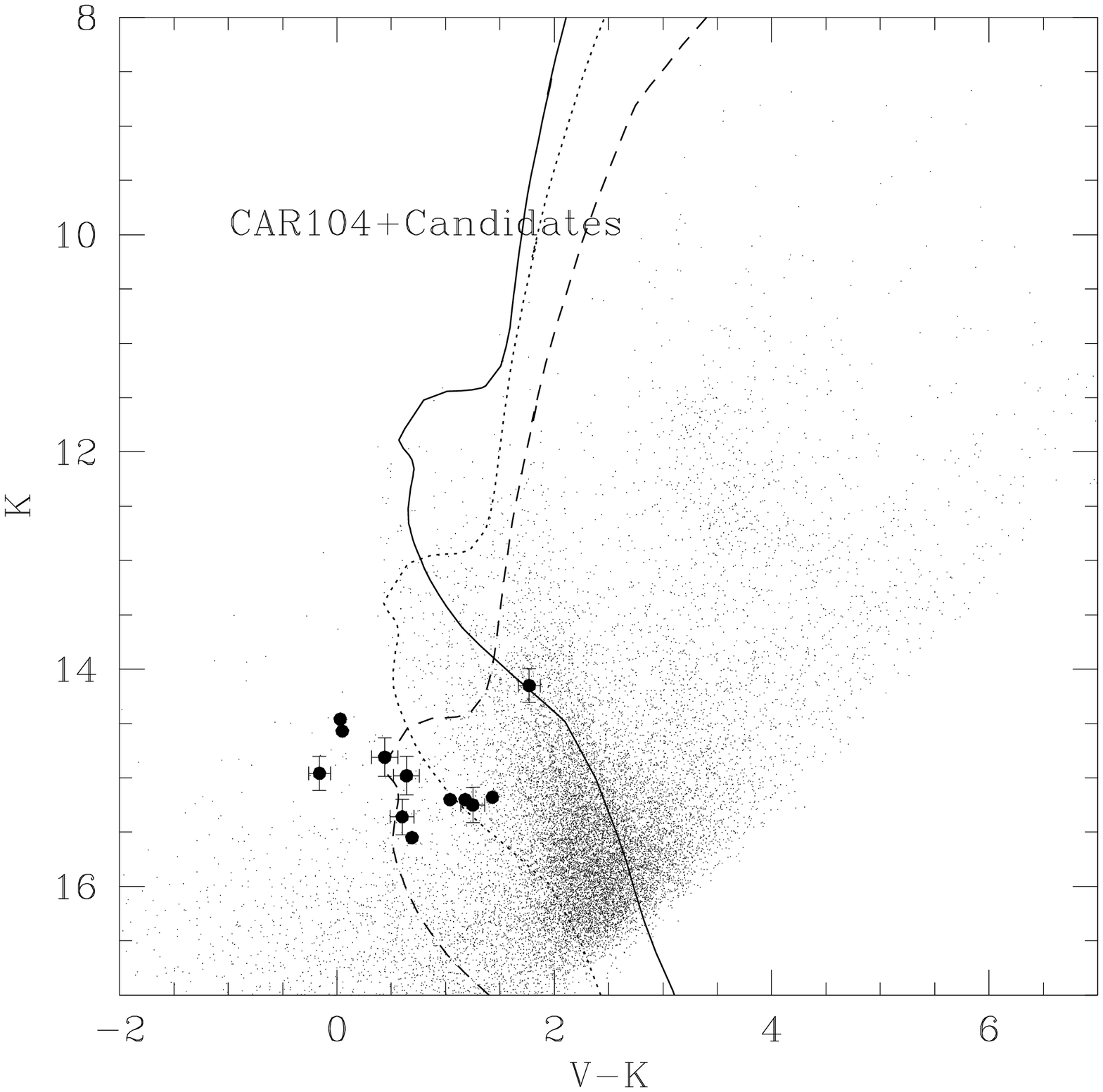}
\includegraphics[height=10cm,width=10cm]{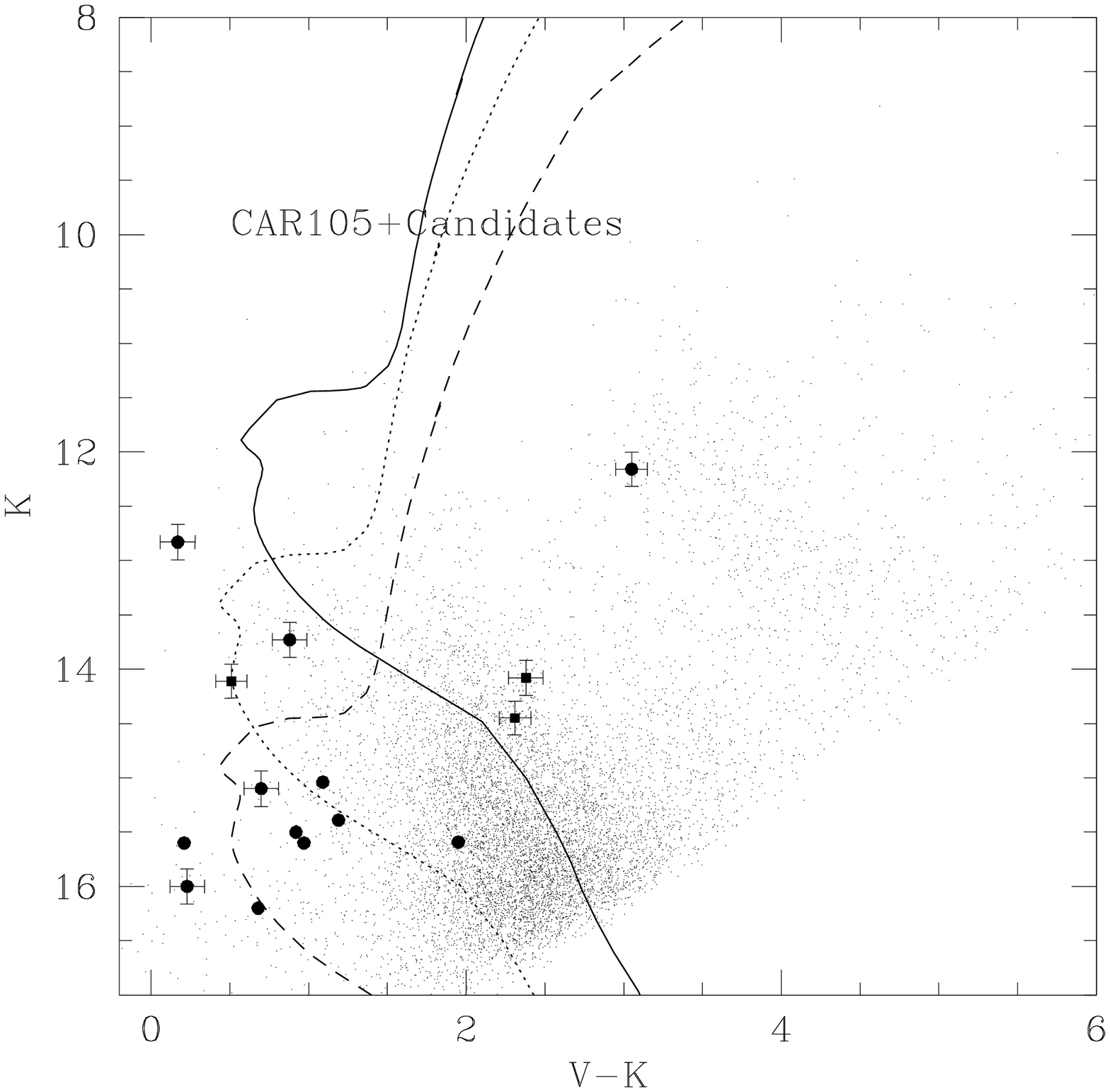}
\caption[]{$K$ vs $V-K$ colour-magnitude diagram for the CAR104 and CAR105 fields. 
The filled circles 
are the transit candidates.  The lines show the Padova isochrones for solar age and 
metallicity (corrected for reddening), at
 distances of 1 kpc (solid line), 2 kpc (dotted line) and 4 kpc (short dashed line). 
Note that, unlike the field stars, stars with transits have been corrected for reddening and
 that the most likely transit candidates to
host exoplanets are marked with full squares.}
\label{fig:KVKcmd}
\end{figure*}

The stellar populations present in the Carina fields not only are interesting in their own right
but could also shed some light on the properties of the transit candidates. For instance, 
in our solar neighborhood many of the confirmed planets orbit around main sequence stars of
perhaps slightly larger metallicity than solar  (Gonzalez 2003). We have seen (\S 4) that the
distances of the transit stars inferred from the SB method range from about 1 to 4 kpc. 
Figures~\ref{fig:IVIcmd}  and ~\ref{fig:KVKcmd} show 
the candidates and the field population in optical colour-magnitude and optical-IR
colour-magnitude diagrams respectively. The Padova isochrones (Girardi et al. 2002) for solar 
age and metallicity, without reddening, are also indicated at 1 kpc (solid line), 2 kpc (dotted line)
and 4 kpc (short dashed line), and confirm that the candidate stars are indeed  within this range. 

Assuming a solar distance to the Galactic center of $R_0=8$ kpc, the location  
of these fields has a typical distance of 7.5 kpc from the Galactic centre.
The stellar population in these fields should not be too different from
that of the solar circle. If we consider the metallicity gradient in the
Galactic disk of -0.09 $\pm$ 0.02 dex/kpc (Friel \& Janes 1993),
the mean metallicity in the field should be $\sim$ 0.04 dex larger in the mean than
that of the Solar neighborhood. Such a small difference would hardly matter.

Further insights can be gained through the comparison with a synthetic model of the Galaxy. 
The Besan\c{c}on model has been  updated recently 
(Robin et al., 2003)\footnote{See {\tt http://www.obs-besancon.fr/www/modele}} and is
very convenient for our purposes as we can derive the distribution of stars, 
their photometric properties and distances in a
given direction. In addition, the model provides information on the metallicities and
kinematics of these samples. The predicted differences between the two fields, Carina 104
and 105 are small given the sampling fluctuations and the patchy extinction.

 From the model, we selected main sequence stars having the same range of  $K$-band magnitudes 
as our candidates. Further, we also selected those with a luminosity class equals to V, that
is, on the main sequence. We also applied a cut in metallicity, such that $-0.25 < \mathrm{[Fe/H]} < 0.5$
but this proves to be not very discriminant, because most stars are within this range, given
the metallicity gradient observed in the disc, as discussed above.
 The question that arises is whether the distributions of inferred 
distances and colour excesses for the candidates could be sampled from the synthetic model. 
This is important as we did not use the  Besan\c{c}on model for the dependence of reddening with
distance, and hence this comparison provides a good test of the procedure, given the model.

\begin{figure*}[h!]
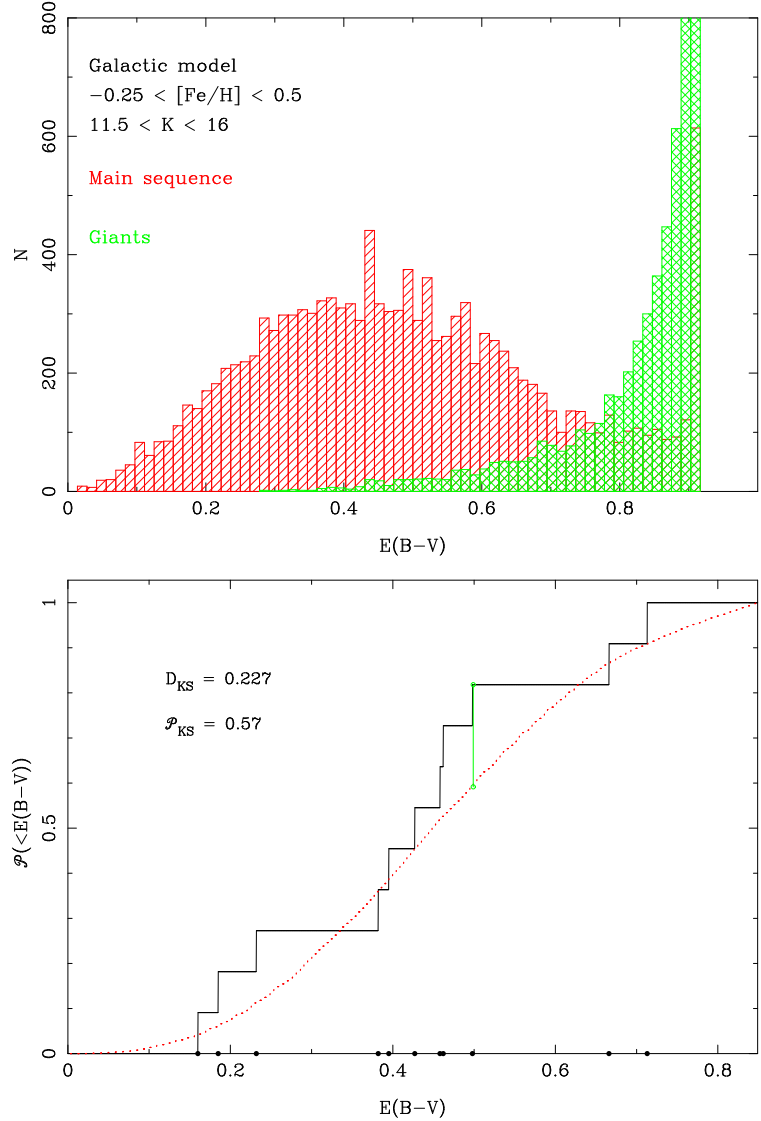

\centering
\includegraphics[height=10cm,angle=-90.]{ebv_besancon.ps}
\includegraphics[height=10cm,angle=-90.]{pks_ebv.ps}
\caption[]{(Top) Distribution function of the \ebv~ colour excess 
predicted by the Besan\c{c}on model for main-sequence
and giant stars in the
range  11.5 $\leq K \leq$ 16.0 and near solar metallicity. Note the gaussian-shaped
distribution for main-sequence stars, and the highly skewed distribution for giants, as
expected for brighter and hence more distant stars. The cutoff at 0.92 corresponds
to the maximal colour excess predicted by the Besan\c{c}on model. 
(Bottom) Cumulative distribution functions of colour excess \ebv~ for the
OGLE transit stars which are inferred to be main-sequence stars (points and dark line) and for
the main sequence stars of the Besan\c{c}on model selected with the criteria of the top panel 
(dotted red line). A two-sample Kolmogorov-Smirnov test gives a 57\% probability.}
\label{fig:ebvmodel}
\end{figure*}

The resulting distribution function of the \ebv~ colour excess is given in Figure~\ref{fig:ebvmodel} (Top)
where for the sake of completeness we have also indicated the distribution for the giant stars. The
gaussian-shaped distribution for the selected main-sequence stars appears very similar to the
distribution of the inferred \ebv~ for the candidates (see Table~\ref{table:VIK}), while giant stars
present a very skewed distribution, a reflection of the brigher absolute magnitudes and hence
larger distances which lead to larger colour excesses, on average. Beyond this qualitative comparison,
we can give a quantitative answer to the question of whether the Besan\c{c}on model could be the
parent population out of which the sample of candidates could have been extracted. Figure~\ref{fig:ebvmodel}
(Bottom) gives the cumulative distribution functions, and a simple two-sample Kolmogorov-Smirnov test
gives a probability of 57\% for the two samples to be drawn from the same population. Note that we
removed the likely giants from our OGLE sub-sample. Although this probability is not too high, we can conclude
that the Besan\c{c}on model provides a reasonable agreement, especially since we did not use the
underlying reddening model. 

\begin{figure*}[h!]
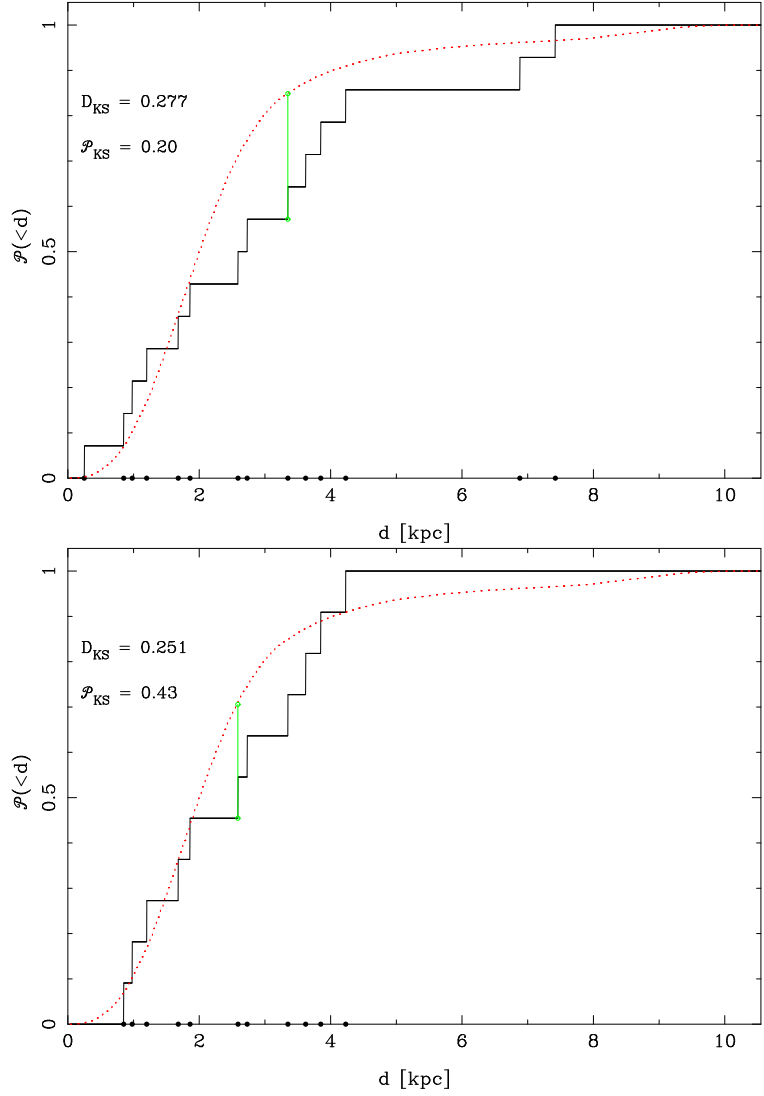

\centering
\includegraphics[height=10cm,angle=-90.]{pks_with102.ps}
\includegraphics[height=10cm,angle=-90.]{pks_nogiants.ps}
\caption[]{Cumulative distribution functions of distances for the
OGLE transit stars with inferred distances (points and dark line) and for
the Besan\c{c}on model with main sequence stars in the 
range 11.5 $\leq K \leq$ 16.0 
(dotted red line). The top panel includes  \ogle{89}, \ogle{102}
and  \ogle{118} and has ${\cal{P}}_{\mathrm{KS}} = 0.20$
while the bottom panel excludes these three giant stars, 
yielding a signficantly much higher probability of ${\cal{P}}_{\mathrm{KS}} = 0.43$.}
\label{fig:galmodel}
\end{figure*}

A further sensitive test comes from the distribution of distances. 
Figure~\ref{fig:galmodel} shows the cumulative distribution functions of both
candidates and synthetic stars (top panel). A Kolmogorov-Smirnov test yields a probability
that the sample of candidates is extracted from the parent synthetic population of 20\% when
we consider all the candidates. 
If we remove \ogle{102}, \ogle{89} and \ogle{118} as likely giant stars,  
we get a much larger probability of 43\%, fully confirming our suspicion that these three stars
cannot be drawn from the main-sequence population.

The detailed distribution of stars in the colour-magnitude diagrams depends heavily on the
star formation history, the metallicity distribution function and the initial stellar mass
 function, in addition to the extinction
properties mentioned in \S 4 and the actual density distributions of the thin and thick discs, 
which are dominant in these fields. While there are methods to infer the past history of star
formation from volume-limited samples close to the solar neighbourhood (e.g. Hernandez et al. 
2000), such an analysis clearly goes beyond the scope of this paper. We will just note here that 
a nearby field ($\ell$ = 292.45, $b$=1.63) has yielded interesting constraints on
galactic structure (see the field F3 analysed by Vallenari et al. 2000) 
although the results depend strongly on the assumed age and metallicity range of the populations. 

\begin{figure*}[h!]
\centering
\includegraphics[width=12cm,angle=-90.]{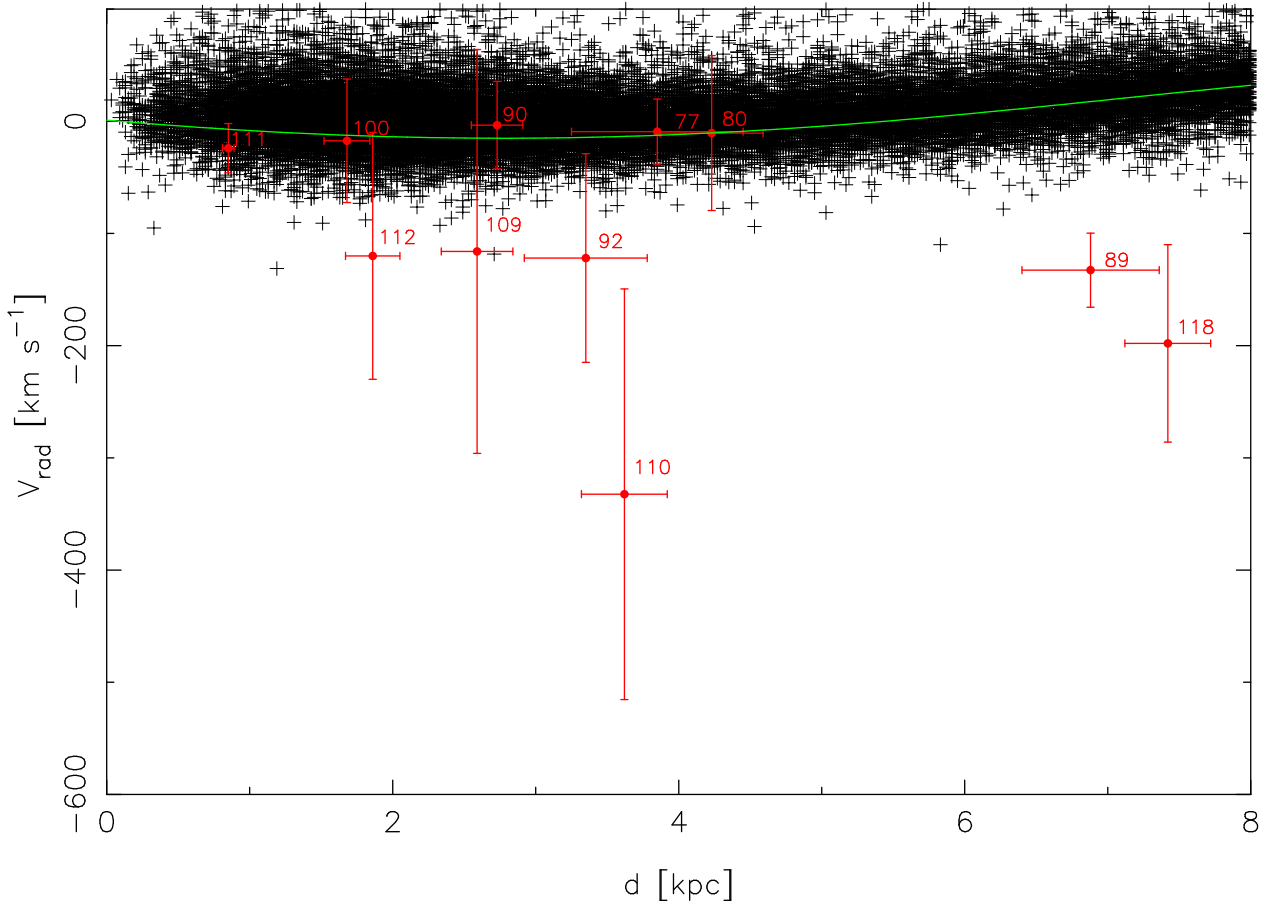}
\caption[]{Radial velocity as a function of distance : transit candidates 
(filled circles with error bars), simulated field
stars from the Besan\c{c}on model (crosses) and mean H{\sc i} 
velocity field (Nakashini \& Sofue 2003). While the majority of OGLE candidates 
are consistent with these models, clearly \ogle{89}, \ogle{110} and \ogle{118} 
cannot be explained within this framework.}
\label{fig:vradgalmodel}
\end{figure*}

Finally, the kinematics of the model can also be tested since we have a few radial velocities
measured from the low-resolution spectra (Table~\ref{table:teffbalmer}). Figure~\ref{fig:vradgalmodel}
gives the radial velocities as a function of the distance along the line
of sight for both the candidates and the synthetic stars. The radial velocity along these
lines of sight can be also calculated using the accurate
model by Nakashini \& Sofue (2003) which fits the observed distribution of H{\sc i} gas.
As expected, most radial velocities are close to zero, and  overall  there 
is an excellent
agreement, except perhaps for \ogle{89}, \ogle{110} and \ogle{118}.

We conclude that, even though these data are not sufficient to strongly constraint the
structure of the Galaxy along these directions, they are consistent with current models.

%--------------------
\section{Conclusions}
%--------------------

In order to make an efficient pre-selection of possible exoplanets, we
obtained $K$-band photometry for a sample of transit candidates detected by the OGLE   
collaboration in the CAR 104 and CAR 105 fields of the Galactic disc,  along with 
low-dispersion spectroscopy of 11 promising  candidates. By
combining these new data with the optical $VI$ photometry from OGLE, we presented a novel
 application of
the surface brightness relations to measure the colour excess of each transit star, its
distance, and, more importantly, the effective temperature and radius, and 
hence to fully 
caracterise their properties. The method rejects automatically giant stars, and provide
accurate solutions for main-sequence stars. A comparison with the temperatures derived 
from the low-resolution spectra shows an excellent agreement and validates the method.
The new measures of stellar radii allow us to further refine  the radii of the companions.

We find that :

1. The combination of optical with near-IR photometry allows a very good estimate the 
stellar parameters for most stars and therefore to select the best planetary transit candidates in a very
effective way, bypassing the need of high-resolution spectroscopy at this stage of the selection
process.

2. We discard \ogle{83}, \ogle{89}, \ogle{98},  \ogle{102}, \ogle{116}, \ogle{118} and \ogle{119}
for being very probably giant stars.

3. Over 70\% of the sub-sample contains low-mass companions whose radii are far too large to
be exoplanets.

4. We select \ogle{105}, \ogle{109} and \ogle{111} (full squares in Fig. 1, 13 and 14) as the most likely transit candidates to
host exoplanets, as the inferred radii of their companions are below  $R_{c} / R_{\odot} < 0.14$.
As their periods are shorter than 3.5 days, these are new possible ``hot Jupiter''-like planets.

5. Recently, Pont et al. (2004) confirmed \ogle{111} as an exoplanet with a mass of 0.53 $\pm$ 0.11 M$_{Jup}$ and a radius of 1.0$^{1.13}_{0.94}$ R$_{Jup}$. Our calculated radius for \ogle{111} is about 0.91 $\pm$ 0.03, showing the reliability of our method to reject ``false positives'' and efficiently select promising extrasolar planetary transit candidates. 
\\
\\
A comparison with the Besan\c{c}on model of the Galaxy shows a satisfactory agreement and that
these stars are normal and representative of the thin disc population.

In the future it would be useful to obtain high-resolution spectra with UVES/VLT for the three 
candidates to measure radial velocities and to determine accurately their masses.\\

\begin{acknowledgements}
We wish to thank the OGLE team, and especially G. Pietrzy\'nski, for making available
the optical $V$ and $I$ magnitudes for the stars used in this study.\\
This work was supported by FONDAP Center for Astrophysics 15010003, 
Foundation ETCHEBES
and by the ECOS/CONICYT program C00U05. 
\end{acknowledgements}

%%%%%%%%%%%%%%%%%%%%%%%%%%%%%%%%%%%%%%%%%%%%%%%%%%%%%%%%%%%%%%%%%%%%%%%%%%%%%%%%%%%%%%%
%
%                                 TABLES
%
%%%%%%%%%%%%%%%%%%%%%%%%%%%%%%%%%%%%%%%%%%%%%%%%%%%%%%%%%%%%%%%%%%%%%%%%%%%%%%%%%%%%%%%

\begin{table*}[h!]
\caption{Coordinates of the observed fields.}
\centering
\begin{tabular}{l c c c c}\hline
\hline 
\multicolumn{1}{c}{Field} & $\alpha$~ (J2000) & $\delta$~ (J2000)& {\it l} &{\sl b} \\ \hline
 CAR~104 & $10{^h}57^{m}30^{s}$ & $-$61\degr40\arcmin00\arcsec&289\fdg84&$-$1\fdg72 \\
 CAR~105 & $10{^h}52^{m}20^{s}$ & $-$61\degr40\arcmin00\arcsec&289\fdg29&$-$1\fdg99 \\ \hline
\end{tabular}
\label{table:fields}
\end{table*}

\begin{table*}[h!]
\caption{Journal of observations.}
\centering
\begin{tabular}{l c c c c}\hline
\hline \multicolumn{1}{c}{Name} & $\alpha$~ (J2000) & $\delta$~ (J2000) & IR Images 
 & Spectra \\ \hline
 \ogle{76} & 10:58:41.90 & $-$61:53:12.3 & NTT+SOFI 18Jan03 & ...\\
 \ogle{77} & 10:58:02.03 & $-$61:49:50.9 & NTT+SOFI 18Jan03 & CLAY+LDSS-2 11May03\\
 \ogle{78} & 10:59:41.62 & $-$61:55:15.0 & NTT+ SOFI 18Jan03 & ...\\
 \ogle{79} & 10:59:35.54 & $-$61:56:59.2 & NTT+ SOFI 18Jan03 & ...\\
 \ogle{80} & 10:57:54.36 & $-$61:42:02.5 & NTT+SOFI 18Jan03 & CLAY+LDSS-2 11May03\\
 \ogle{83} & 10:57:42.48 & $-$61:36:23.3 & NTT+SOFI 18Jan03 & ...\\
 \ogle{84} & 10:59:00.00 & $-$61:34:43.0 & NTT+SOFI 18Jan03 & ...\\
 \ogle{85} & 10:59:00.18 & $-$61:37:41.1 & NTT+SOFI 18Jan03 & ...\\
 \ogle{87} & 10:59:39.33 & $-$61:24:07.3 & NTT+SOFI 18Jan03 & ...\\
 \ogle{89} & 10:56:11.27 & $-$61:29:55.4 & NTT+SOFI 18Jan03 & CLAY+LDSS-2 10May03\\
 \ogle{90} & 10:56:36.63 & $-$61:28:46.5 & NTT+SOFI 18Jan03 & CLAY+LDSS-2 10May03\\
 \ogle{92} & 10:57:23.43 & $-$61:26:45.4 & NTT+SOFI 18Jan03 & CLAY+LDSS-2 11May03\\
 \ogle{95} & 10:55:19.38 & $-$61:32:12.0 & NTT+SOFI 18Jan03 & ...\\
 \ogle{98} & 10:56:51.77 & $-$61:56:15.0 & NTT+SOFI 18Jan03 & ...\\
 \ogle{99} & 10:55:12.80 & $-$61:54:54.8 & NTT+SOFI 18Jan03 & ...\\ \hline
 \ogle{100} & 10:52:56.91 & $-$61:50:54.9 & NTT+SOFI 18Jan03 & CLAY+LDSS-2 10May03\\
 \ogle{101} & 10:52:58.59 & $-$61:51:43.1 & NTT+SOFI 18Jan03 & CLAY+LDSS-2 11May03\\
 \ogle{102} & 10:53:29.65 & $-$61:47:37.2 & NTT+SOFI 18Jan03 & ...\\
 \ogle{104} & 10:53:27.04 & $-$61:43:20.3 & NTT+SOFI 18Jan03 & ...\\
 \ogle{105} & 10:52:24.07 & $-$61:31:09.4 & NTT+SOFI 18Jan03 & ...\\
 \ogle{106} & 10:53:51.23 & $-$61:34:13.2 & NTT+SOFI 18Jan03 & ...\\
 \ogle{107} & 10:54:23.58 & $-$61:37:21.1 & NTT+SOFI 18Jan03 & ...\\
 \ogle{108} & 10:53:12.65 & $-$61:30:18.7 & NTT+SOFI 18Jan03 & ...\\
 \ogle{109} & 10:53:40.73 & $-$61:25:14.8 & NTT+SOFI 18Jan03 & CLAY+LDSS-2 11May03\\
 \ogle{110} & 10:52:28.37 & $-$61:29:31.8 & NTT+SOFI 18Jan03 & CLAY+LDSS-2 11May03\\
 \ogle{111} & 10:53:17.91 & $-$61:24:20.3 & NTT+SOFI 18Jan03 & CLAY+LDSS-2 10May03\\
 \ogle{112} & 10:52:46.46 & $-$61:23:17.7 & NTT+SOFI 18Jan03 & CLAY+LDSS-2 11May03\\
 \ogle{116} & 10:50:24.79 & $-$61:26:12.2 & NTT+SOFI 18Jan03 & ...\\
 \ogle{117} & 10:51:40.48 & $-$61:34:15.7 & NTT+SOFI 18Jan03 & ...\\
 \ogle{118} & 10:51:32.10 & $-$61:48:08.3 & NTT+SOFI 18Jan03 & CLAY+LDSS-2 11May03\\
 \ogle{119} & 10:51:58.75 & $-$61:41:20.5 & NTT+SOFI 18Jan03 & ...\\
 \ogle{120} & 10:51:09.34 & $-$61:43:11.3 & NTT+SOFI 18Jan03 & ...\\ \hline
\end{tabular}
\label{table:logobs}
\end{table*}

\begin{table*}[h] 
\caption{Observed and derived photometric properties} 
\begin{center}
\begin{tabular}{lcccccccc}\hline 
 \hline
\multicolumn{1}{c}{  Name}      &  V$^a$ & I$^a$ &  K$^b$                  &   E(B--V)          & $V_o$             & $(V-I)_o$         & $(V-K)_o$ &  $M_V$ \\ \hline
\ogle{76}  & 14.30  & 13.76 & 13.48 $\pm$  0.01   &                   &                   &                   &                   &        \\ 
\ogle{77}  & 17.46  & 16.12 & 15.06 $\pm$  0.03   &  0.71 $\pm$  0.05 & 15.25 $\pm$  0.13 &  0.19 $\pm$  0.15 &  0.44 $\pm$  0.12 &  2.32 \\ 
\ogle{78}  & 16.23 & 15.32 & 14.75 $\pm$  0.02    &  0.52:            & 14.62:            &  0.07:            &  0.05:            &  1.34:  \\ 
\ogle{79}  & 16.04 & 15.28 & 14.64 $\pm$  0.03 &  0.50: & 14.49: & -0.05: &  0.03: &  1.39:  \\ 
\ogle{80}  & 17.50 & 16.50 & 15.53 $\pm$  0.04 &  0.50 $\pm$ 0.02 & 15.96 $\pm$  0.12 &  0.20 $\pm$  0.14 &  0.60 $\pm$  0.11 &  2.82 \\ 
\ogle{83}  & 15.46 & 14.87 & 14.38 $\pm$  0.01   &   &  &  &   & \\ 
\ogle{84}  & 17.79 & 16.69 & 15.73 $\pm$  0.05 &  0.50: & 16.24: &  0.29: &  0.69: &  3.13:  \\ 
\ogle{85}  & 16.64 & 15.45 & 14.23 $\pm$  0.02 &  0.23 $\pm$  0.01 & 15.92 $\pm$  0.12 &  0.82 $\pm$  0.14 &  1.77 $\pm$  0.10 &  5.52 \\ 
\ogle{87}  & 17.51 & 16.32 & 15.34 $\pm$  0.04 &  0.41: & 16.24: &  0.53: &  1.04: &  3.93:  \\ 
\ogle{89}  & 16.63 & 15.78 & 15.17 $\pm$  0.03 &  0.59 $\pm$  0.01 & 14.80 $\pm$  0.12 & -0.10 $\pm$  0.14 & -0.16 $\pm$  0.10 &  0.62 \\ 
\ogle{90}  & 17.72 & 16.44 & 15.38 $\pm$  0.03 &  0.40 $\pm$  0.02 & 16.50 $\pm$  0.12 &  0.64 $\pm$  0.14 &  1.25 $\pm$  0.11 &  4.31 \\ 
\ogle{92}  & 17.68 & 16.50 & 15.21 $\pm$  0.04 &  0.67 $\pm$  0.05 & 15.62 $\pm$  0.13 &  0.10 $\pm$  0.15 &  0.64 $\pm$  0.12 &  2.99 \\ 
\ogle{95}  & 17.73 & 16.36 & 15.31 $\pm$  0.05 &  0.36: & 16.61: &  0.79: &  1.43: &  4.79:  \\ 
\ogle{98}  & 17.46 & 16.64 & 16.14 $\pm$  0.05   &   &  &  &   & \\ 
\ogle{99}  & 17.56 & 16.47 & 15.34 $\pm$  0.04 &  0.38: & 16.38: &  0.48: &  1.18: &  4.32:  \\ \hline
\ogle{100} & 15.93 & 14.88 & 13.88 $\pm$  0.02 &  0.43 $\pm$  0.03 & 14.61 $\pm$  0.12 &  0.36 $\pm$  0.14 &  0.88 $\pm$  0.11 &  3.48 \\ 
\ogle{101} & 17.85 & 16.69 & 15.67 $\pm$  0.04 &  0.46: & 16.42: &  0.42: &  0.92: &  3.67:  \\ 
\ogle{102} & 15.33 & 13.84 & 12.17 $\pm$  0.02 &  0.04 $\pm$  0.002 & 15.21 $\pm$  0.12 &  1.43 $\pm$  0.14 &  3.05 $\pm$  0.10 &  8.22 \\ 
\ogle{104} & 18.53 & 17.10 & 15.70 $\pm$  0.03 &  0.32: & 17.54: &  0.91: &  1.95: &  5.99:  \\ 
\ogle{105} & 17.33 & 16.16 & 14.51 $\pm$  0.01 &  0.19 $\pm$  0.01 & 16.76 $\pm$  0.12 &  0.87 $\pm$  0.14 &  2.31 $\pm$  0.10 &  6.80 \\ 
\ogle{106} & 17.85 & 16.53 & 15.53 $\pm$  0.04 &  0.41: & 16.58: &  0.66: &  1.19: &  4.22:  \\ 
\ogle{107} & 17.58 & 16.66 & 15.80 $\pm$  0.02 &  0.57: & 15.81: &  0.00: &  0.21: &  1.96:  \\ 
\ogle{108} & 18.65 & 17.28 & 16.40 $\pm$  0.05 &  0.57: & 16.88: &  0.45: &  0.68: &  2.97:  \\ 
\ogle{109} & 15.80 & 14.99 & 14.24 $\pm$  0.02 &  0.38 $\pm$  0.02 & 14.62 $\pm$  0.12 &  0.19 $\pm$  0.14 &  0.51 $\pm$  0.10 &  2.55 \\ 
\ogle{110} & 17.23 & 16.15 & 15.26 $\pm$  0.03 &  0.46 $\pm$  0.02 & 15.80 $\pm$  0.12 &  0.33 $\pm$  0.14 &  0.70 $\pm$  0.11 &  3.00 \\ 
\ogle{111} & 16.96 & 15.55 & 14.14 $\pm$  0.04 &  0.16 $\pm$  0.01 & 16.46 $\pm$  0.12 &  1.15 $\pm$  0.14 &  2.38 $\pm$  0.11 &  6.82 \\ 
\ogle{112} & 14.42 & 13.64 & 12.99 $\pm$  0.02 &  0.46 $\pm$  0.03 & 13.00 $\pm$  0.12 &  0.04 $\pm$  0.15 &  0.17 $\pm$  0.11 &  1.65 \\ 
\ogle{116} & 15.47 & 14.90 & 14.47 $\pm$  0.01   &   &  &  &   & \\ 
\ogle{117} & 18.03 & 16.71 & 15.77 $\pm$  0.04 &  0.47: & 16.57: &  0.56: &  0.97: &  3.69:  \\ 
\ogle{118} & 18.09 & 17.07 & 16.21 $\pm$  0.05 &  0.60 $\pm$  0.005 & 16.23 $\pm$  0.12 &  0.05 $\pm$  0.14 &  0.23 $\pm$  0.11 &  1.88 \\ 
\ogle{119} & 14.75 & 14.29 & 13.80 $\pm$  0.02   &   &  &  &   &  \\ 
\ogle{120} & 17.31 & 16.23 & 15.18 $\pm$  0.05 &  0.38: & 16.13: &  0.47: &  1.09: &  4.09:  \\ 
 \hline 
\end{tabular}
\end{center}
\noindent $^a$ The uncertainties in the $V$ and $I$ magnitudes are of order 0.10--0.15 magnitudes 
(Udalski et al. 2002a, 2002b).

\noindent $^b$ Errors do not include the zero-point uncertainty.
\label{table:VIK}
\end{table*}

\begin{table*}[h!] 
\caption{Effective temperatures derived from the Balmer lines and radial velocities.} 
\centering
\begin{tabular}{lrrrrrrrr}\hline 
 \hline
\multicolumn{1}{c}{Name} &    
\multicolumn{1}{c}{EW(H$_{\gamma}$)} &
\multicolumn{1}{c}{EW(H$_{\beta}$)}  & 
\multicolumn{1}{c}{EW(H$_{\alpha}$)} & 
\multicolumn{1}{c}{T$_{\mathrm{eff}}$(H$_{\gamma}$)} &   
\multicolumn{1}{c}{T$_{\mathrm{eff}}$(H$_{\beta}$)}  &
\multicolumn{1}{c}{T$_{\mathrm{eff}}$(H$_{\alpha}$)} &
\multicolumn{1}{c}{T$_{\mathrm{eff}}$(Balmer)}    & 
\multicolumn{1}{c}{v$_{\mathrm{rad}}$}\\
       &  \multicolumn{1}{c}{[\AA]}       &  \multicolumn{1}{c}{[\AA]}       & \multicolumn{1}{c}{[\AA]}      &
\multicolumn{1}{c}{[K]}  &  \multicolumn{1}{c}{[K]}  &  \multicolumn{1}{c}{[K]}  &
  \multicolumn{1}{c}{[K]} &   \multicolumn{1}{c}{[km/s]} \\ \hline 
\ogle{77} &   6.7  &     4.7  &  5.2: & 7460 $\pm$ 600 &     7000 $\pm$ 600  &     7600 $\pm$ 255  & 7300 $\pm$ 480 & --9 $\pm$ 30 \\
\ogle{80} &   5.6  &     5.8  &  4.7  & 7200 $\pm$ 700 &     7400 $\pm$ 600  &     7470 $\pm$ 900  & 7360 $\pm$ 730 & --10 $\pm$ 70  \\
\ogle{89} &   1.6  &     2.3  &  3.1  & --           &     5700 $\pm$ 1300 &     6430 $\pm$ 1000 & 6030 $\pm$ 1100  & --130 $\pm$ 30 \\
\ogle{90} &   1.0  &     2.7  &  1.3: &  --          &     6200 $\pm$ 1200 &     4620 $\pm$ 1000 & 5470 $\pm$ 1100  & --3 $\pm$ 40 \\
\ogle{92} &   --   &     4.4  &  3.6  & --           &     6900 $\pm$ 750  &     6820 $\pm$ 1000 & 6860 $\pm$ 900   & --120 $\pm$ 90 \\ \hline
\ogle{100} &  6.0  &     5.2  &  4.1   & 7300 $\pm$ 600&     7160 $\pm$ 600  &     7100 $\pm$ 1000 & 7200 $\pm$ 700 & --17 $\pm$ 50  \\
\ogle{109} &  7.7  &     10.0 &  6.9   & 7700 $\pm$ 600&     8360 $\pm$ 600  &     8100 $\pm$ 1000 & 8060 $\pm$ 700 & --120 $\pm$ 180  \\
\ogle{110} &  3.5  &     3.9  &  5.1  &  6200 $\pm$ 600&     6750 $\pm$ 1500 &     7600 $\pm$ 750  & 6850 $\pm$ 950 & --330 $\pm$ 180\\
\ogle{111} &  --   &       -- &  1.0  &   --         &          --       &     4460 $\pm$ 1000 & 4460 $\pm$ 1000    & --24 $\pm$ 20  \\
\ogle{112} &  10.9 &     11.0 &  8.1  &  8600 $\pm$ 600&     8960 $\pm$ 600  &     9000 $\pm$ 330  & 8830 $\pm$ 500 & --120 $\pm$ 110 \\
\ogle{118} &  5.2  &      6.8 &  3.0  &  8650 $\pm$ 450&     8570 $\pm$ 600  &     9000 $\pm$ 300  & 8690 $\pm$ 450 & --200 $\pm$ 90\\ 
\hline
\end{tabular}
\label{table:teffbalmer}
\end{table*}

\begin{table*}[h!] 
\caption{Inferred properties for the stars and their companions.} 
\begin{center}
\begin{tabular}{lrrrrrrr}\hline 
 \hline
\multicolumn{1}{c}{  Name } &    \multicolumn{1}{c}{$d$}   &  \multicolumn{1}{c}{E(B-V)} &  \multicolumn{1}{c}{T$_{\mathrm{eff}}$(SB)}& 
  \multicolumn{1}{c}{$R_*$(SB)} & \multicolumn{1}{c}{$\Delta I$}  &   \multicolumn{1}{c}{$R_c$} &
\multicolumn{1}{c}{$P$} \\
       &    \multicolumn{1}{c}{[kpc]} &  \multicolumn{1}{c}{[mag]}       &     \multicolumn{1}{c}{[K]}           &
   \multicolumn{1}{c}{[$R_{\odot}$]} &  \multicolumn{1}{c}{[mag]} & \multicolumn{1}{c}{[$R_{\odot}$]} & \multicolumn{1}{c}{[days]} \\ \hline 
\ogle{77}   &    3.85  $\pm$ 0.60 &0.713 $\pm$ 0.054 &  7850  $\pm$  640 &    1.61  $\pm$  0.20 &
   0.022  & 0.230 $\pm$ 0.029 & 5.45550 \\ 
\ogle{78}   &     4.53 :          & 0.52 :           &   9000 :          &    2.29 :            & 
            &                     &     \\
\ogle{79}   &     4.16 :          & 0.50 :           &   8900 :          &    2.27 :            &
            &                     &     \\
\ogle{80}   &    4.23  $\pm$ 0.36 &0.498 $\pm$ 0.019 &  7280  $\pm$  300 &    1.43  $\pm$  0.10 &
  0.016 &  0.174 $\pm$ 0.012 & 1.80730 \\
\ogle{83}   & \multicolumn{4}{c}{probably giant }                                            &
           &                      &    \\
\ogle{84}   &   4.2 :            & 0.50 :           &  6900 :           &   1.4 :         &
            &              &   \\
\ogle{85}   &    1.20  $\pm$ 0.07 &0.232 $\pm$ 0.015 &  5300  $\pm$  140 &    0.87  $\pm$  0.04 &
    0.048 & 0.185 $\pm$ 0.009 &  2.11460  \\
\ogle{87}   &  2.9 -- 7.8 :     &0.4 -- 0.9 :    & 6300 -- 11000 :   & 1.2 -- 2.6 :       &
       &           &      \\
\ogle{89}$^a$   &    6.88  $\pm$ 0.48 &0.589 $\pm$ 0.010 & 10200  $\pm$  400 &    2.42  $\pm$  0.14 &
     0.013 & 0.266 $\pm$ 0.015 & 2.28990 \\ 
\ogle{90}   &    2.73  $\pm$ 0.18 &0.395 $\pm$ 0.016 &  6090  $\pm$  200 &    1.08  $\pm$  0.05 &
          0.022 & 0.155 $\pm$ 0.007  & 1.04155 \\ 
\ogle{92}   &    3.35  $\pm$ 0.43 &0.666 $\pm$ 0.046 &  7090  $\pm$  470 &    1.37  $\pm$  0.14 &
       0.038 & 0.259 $\pm$ 0.026 & 0.97810 \\ 
\ogle{95}   &  2.3 -- 6.0 :       & 0.4 -- 0.9 :     & 5700 -- 9400 :    &  1.0 -- 2.1 :     &
         &         &     \\
\ogle{98}   & \multicolumn{4}{c}{probably giant }                                 &
        &          &            \\
\ogle{99}   &  2.6 -- 5.6 :       & 0.4 -- 0.9 :     & 6000 -- 9700 :    & 1.1 -- 2.2 :       &
        &          &           \\ \hline
\ogle{100}  &    1.68  $\pm$ 0.16 &0.427 $\pm$ 0.030 &  6700  $\pm$  320 &    1.26  $\pm$  0.09 &
   0.019 & 0.167 $\pm$ 0.012 & 0.82670  \\ 
\ogle{101}  &  3.6 :              & 0.46 :           & 6500 :            &    1.3 :            &
        &                &        \\
\ogle{102}$^b$  &    0.25  $\pm$ 0.01 &0.039 $\pm$ 0.002 &  4040  $\pm$   60 &    0.57  $\pm$  0.01 & 
    0.019 &  0.076 $\pm$ 0.001 & 3.09790 \\
\ogle{104}  &  2.0 -- 3.7 :       & 0.3 -- 0.7 :     & 5000 -- 6700 :    & 0.8 -- 1.3 :   &
          &        &      \\      
\ogle{105}  &    0.98  $\pm$ 0.05 &0.185 $\pm$ 0.010 &  4600  $\pm$  100 &    0.70  $\pm$  0.02 &
         0.026 &  0.109 $\pm$ 0.003 & 3.05810 \\
\ogle{106}  &  2.9 -- 6.9 :       & 0.4 -- 0.9 :     & 6100 -- 9600 :    & 1.1 -- 2.2  :  &
       &        &         \\
\ogle{107}  & 5.9 :               & 0.57 :           & 8100 :            & 1.8 :          &
        &       &        \\
\ogle{108}  & 6.0 :               & 0.57 :           & 7200 :            & 1.5 :          & 
       &         &       \\   
\ogle{109}  &    2.59  $\pm$ 0.25 &0.382 $\pm$ 0.023 &  7580  $\pm$  370 &    1.52  $\pm$  0.12 &
  0.008 & 0.131 $\pm$ 0.010 & 0.58909 \\
\ogle{110}  &    3.62  $\pm$ 0.30 &0.462 $\pm$ 0.020 &  7160  $\pm$  300 &    1.39  $\pm$  0.09 & 
    0.026 & 0.216 $\pm$ 0.014  & 2.84857              \\ 
\ogle{111}  &    0.85  $\pm$ 0.04 &0.160 $\pm$ 0.009 &  4650  $\pm$   95 &    0.71  $\pm$  0.02 &
    0.019 & 0.094 $\pm$ 0.003 & 4.01610 \\  
\ogle{112}  &    1.86  $\pm$ 0.19 &0.458 $\pm$ 0.034 &  8600  $\pm$  470 &    1.85  $\pm$  0.16 &
     0.016 & 0.225 $\pm$ 0.019  & 3.87900   \\ 
\ogle{116}  & \multicolumn{4}{c}{probably giant }                                       &
        &            &   \\
\ogle{117}  &  3.8 :              & 0.47 :           & 6500 :            &  1.3 :        &
         &           &  \\
\ogle{118}$^a$  &    7.42  $\pm$ 0.30 &0.600 $\pm$ 0.005 &  8300  $\pm$  170 &    1.75  $\pm$  0.06 & 
    0.019 & 0.233 $\pm$ 0.008 & 1.86150  \\ 
\ogle{119}  & \multicolumn{4}{c}{probably giant }                                   &
          &                   &         \\
\ogle{120}  &  2.6 -- 6.3 :       & 0.4 -- 0.9 :     & 6100 -- 10000 :   & 1.2 -- 2.3 :  &
           &              &        \\
\hline
\end{tabular}
\end{center}

\noindent $^a$ The large distance inferred from the SB method indicates that this star is likely to be a giant.

\noindent $^b$ Although this is, formally, a main-sequence solution, this is star is likely to be a giant.

\label{table:sbfinal}
\end{table*}

\clearpage

%%%%%%%%%%%%%%%%%%%%%%%%%%%%%%%%%%%%%%%%%%%%%%%%%%%%%%%%%%%%%%%%%%%%%%%%%%%%%%%%%%%%%%
%
%                                       FIGURES
%
%%%%%%%%%%%%%%%%%%%%%%%%%%%%%%%%%%%%%%%%%%%%%%%%%%%%%%%%%%%%%%%%%%%%%%%%%%%%%%%%%%%%%%

\clearpage

\clearpage

\clearpage

\end{document}